\documentclass[fleqn,10pt]{wlscirep}
\usepackage[utf8]{inputenc}
\usepackage[T1]{fontenc}
\usepackage{graphicx,epsf,bm,amsmath,amsfonts,amssymb,epstopdf,hyperref,color,verbatim,multirow,bm}
\hypersetup{colorlinks=true,urlcolor=blue,citecolor=blue,linkcolor=blue,menucolor=blue,anchorcolor=blue,filecolor=blue}

\title{Event Horizon Telescope observations exclude compact objects in baseline mimetic gravity}

\author[1]{Mohsen Khodadi}
\author[2,3*]{Sunny Vagnozzi}
\author[1,4,5]{Javad T. Firouzjaee}
\affil[1]{Department of Physics, K. N. Toosi University of Technology, P. O. Box 15875-4416, Tehran, Iran}
\affil[2]{Department of Physics, University of Trento, Via Sommarive 14, 38123 Povo (TN), Italy}
\affil[3]{Trento Institute for Fundamental Physics and Applications-INFN, Via Sommarive 14, 38123 Povo (TN), Italy}
\affil[4]{PDAT Laboratory, Department of Physics, K. N. Toosi University of Technology, P. O. Box 15875-4416, Tehran, Iran}
\affil[5]{School of Physics, Institute for Research in Fundamental Sciences (IPM), P. O. Box 19395-5531, Tehran, Iran}

\affil[*]{sunny.vagnozzi@unitn.it}


\begin{abstract}
Mimetic gravity has gained significant appeal in cosmological contexts, but static spherically symmetric space-times within the baseline theory are highly non-trivial: the two natural solutions are a naked singularity and a black hole space-time obtained through an appropriate gluing procedure. We study the shadow properties of these two objects, finding both to be pathological. In particular, the naked singularity does not cast a shadow, whereas the black hole casts a shadow which is too small. We argue that the Event Horizon Telescope images of M87$^{\star}$ and Sgr A$^{\star}$ rule out the baseline version of mimetic gravity, preventing the theory from successfully accounting for the dark sector on cosmological scales. Our results highlight an interesting complementarity between black hole imaging observations and modified gravity theories of cosmological interest.
\end{abstract}

\begin{document}

\flushbottom
\maketitle

\thispagestyle{empty}

\section*{Introduction}
\label{sec:introduction}

Black holes (BHs) are arguably among the most peculiar regions of the Universe. While they were once considered to be objects of mere mathematical interest, groundbreaking advances in various fields have made it possible to observe on a regular basis a wide range of astrophysical effects associated to BHs, whose existence is now undisputed. These advances have turned BHs into the ideal testbed for strong-field tests of fundamental physics, including the nature of the dark matter (DM) and dark energy (DE) which pervade the Universe: of interest to us is the possibility that these components may be the manifestation of a theory of gravity beyond General Relativity (GR)~\cite{Nojiri:2010wj,Clifton:2011jh,Nojiri:2017ncd}, and mimetic gravity is a particularly interesting theory in this sense.

Mimetic gravity provides a framework where the conformal mode of gravity is isolated covariantly by means of a scalar field, the mimetic field $\phi$~\cite{Chamseddine:2013kea,Sebastiani:2016ras}. This procedure leads to the appearance of an additional (conformal) degree of freedom. On cosmological scales, the conformal mode in mimetic gravity becomes dynamical even in the absence of matter and behaves as pressureless dust, thereby effectively mimicking DM. On the other hand, relatively simple extensions of mimetic gravity, several of which have been studied in recent years~\cite{Chamseddine:2014vna,Chaichian:2014qba,Nojiri:2014zqa,Mirzagholi:2014ifa,Leon:2014yua,Momeni:2015gka,Myrzakulov:2015qaa,Astashenok:2015haa,Rabochaya:2015haa,Odintsov:2016oyz,Oikonomou:2016pkp,Firouzjahi:2017txv,Hirano:2017zox,Zheng:2017qfs,Vagnozzi:2017ilo,Takahashi:2017pje,Gorji:2017cai,Dutta:2017fjw,Odintsov:2018ggm,Casalino:2018tcd,Ganz:2018mqi,Solomon:2019qgf,Gorji:2019ttx,Khalifeh:2019zfi,Rashidi:2020jao,Kaczmarek:2021psy,Benisty:2021cin,Nashed:2022yfc,Nashed:2023jdf,Kaczmarek:2023qmq}, can easily mimic arbitrary cosmological backgrounds, including ones featuring both DM and DE. Mimetic gravity has therefore gained significant interest over the past decade as a candidate modified gravity model for the dark sector (at least at the background level), and more generally as a testbed for the possible role of disformal transformations in cosmology. In spite of all the interest it received, mimetic gravity has undergone virtually no direct tests against observations (one of the few robust observational tests of mimetic gravity was performed against the multi-messenger gravitational wave event GW170817~\cite{Casalino:2018wnc}), and our goal is to to fill this gap with a highly non-trivial test of the theory against horizon-scale BH observations from the Event Horizon Telescope (EHT).

As we shall discuss in more detail later, the nature of compact objects in mimetic gravity is highly non-trivial. In fact, unless one is willing to accept an imaginary field, the mimetic gravity scenario of cosmological interest where $\partial_{\mu}\phi$ is timelike supports no natural BH solutions, which is already pathological per se given the deluge of observational evidence for the existence of astrophysical BHs. Later work showed that baseline mimetic gravity allows for more model freedom than previously thought, in relation to the sign controlling the underlying conformal transformation~\cite{Gorji:2020ten}: nevertheless, it still remains highly non-trivial to construct consistent solutions for compact objects in mimetic gravity. Essentially two classes of compact object solutions exist in baseline mimetic gravity~\cite{Gorji:2020ten}: naked singularities (NSs), and BHs obtained through a peculiar ``gluing'' procedure. Neither of these space-times are trivial, so the question naturally arises of whether compact objects described by these metrics are consistent with current BH-related observations: as we shall see at the end of this work, the answer is no.

Our goal in this work is to assess the observational consistency of mimetic compact objects, focusing on the horizon-scale images of the supermassive BHs (SMBHs) M87$^{\star}$ and Sgr A$^{\star}$ provided by the EHT in 2019 and 2022 respectively~\cite{EventHorizonTelescope:2019dse,EventHorizonTelescope:2022wkp}. To do so, we will study the shadow properties of mimetic compact objects (which are of direct relevance in assessing the optical appearance thereof), showing that these properties are pathological for both the mimetic NS and mimetic BH. In particular, the former is found to not cast a shadow, whereas the latter casts a shadow which is too small. Both behaviors allow us to conclude that mimetic compact objects can in no way be consistent with the EHT observations. As these two are the only natural compact object solutions in baseline mimetic gravity, we conclude that the EHT observations rule out the basic version of mimetic gravity, which is excluded from playing an important role in cosmology.

The rest of this paper is then organized as follows. We begin by briefly reviewing the key features of mimetic gravity compact objects therein, introducing the mimetic NS and mimetic BH. The shadow properties of these two space-times are then studied and argued to be pathological. We then draw concluding remarks. Throughout the work, we use natural units where $M_{\text{Pl}}=1/\sqrt{8\pi{G}}=1$, and adopt the ``mostly plus'' metric signature $(-,+,+,+)$.

\section*{Mimetic gravity and mimetic compact objects}
\label{sec:mimetic}

In mimetic gravity, the conformal mode of gravity is isolated covariantly by re-parametrizing the physical metric $g_{\mu\nu}$ in terms of an auxiliary metric $\tilde{g}_{\mu\nu}$ and the mimetic scalar field $\phi$~\cite{Chamseddine:2013kea,Sebastiani:2016ras}:
\begin{eqnarray}
g_{\mu\nu} = \pm \big( {\tilde g}^{\alpha\beta}\partial_{\alpha}\phi\partial_{\beta}\phi \big)\, {\tilde g}_{\mu\nu} \,.
\label{eq:mimetic}
\end{eqnarray}
The physical metric is clearly invariant under conformal transformations of the auxiliary metric, with the following required to hold for consistency:
\begin{eqnarray}
g^{\mu\nu} \partial_\mu \phi \partial_\nu \phi  = \pm 1 \,.
\label{eq:mimeticconstraint}
\end{eqnarray}
With our choice of metric signature, $\partial_{\mu}\phi$ is timelike (spacelike) when the sign in Eqs.~(\ref{eq:mimetic},\ref{eq:mimeticconstraint}) is taken to be the negative (positive) one. Varying the Einstein-Hilbert action with respect to the physical metric, taking into account its dependence on $\tilde{g}_{\mu\nu}$ and $\phi$ as in Eq.~(\ref{eq:mimetic}), leads to an additional contribution which mimics cosmological (cold) DM. The appearance of the additional degree of freedom can be understood in terms of the transformation between the physical and auxiliary metric being a singular (non-invertible) disformal transformation~\cite{Zumalacarregui:2013pma,Deruelle:2014zza,Domenech:2015tca,Arroja:2015wpa,BenAchour:2016cay,Jirousek:2022rym,Domenech:2023ryc}, which therefore renders the longitudinal mode of gravity dynamical even in the absence of matter fields.

The mimetic constraint Eq.~(\ref{eq:mimeticconstraint}) can be introduced at the level of the action via a Lagrange multiplier term~\cite{Golovnev:2013jxa}:
\begin{eqnarray}
S_{\pm}= \int d^4 x  \sqrt{-g} \left[\, \frac{R}{2} + \lambda \Big( g^{\mu \nu } \partial_\mu \phi \partial_\nu \phi \mp 1 \Big) \right] \,,
\label{eq:action}
\end{eqnarray}
where $R$ is the Ricci scalar constructed from the physical metric $g_{\mu\nu}$, while the auxiliary field $\lambda$ enforces the mimetic constraint Eq.~(\ref{eq:mimeticconstraint}), and the $S_-$ ($S_+$) action corresponds to timelike (spacelike) $\partial_{\mu}\phi$. The above formulation also makes the connection to related earlier models clearer~\cite{Lim:2010yk,Gao:2010gj,Capozziello:2010uv}. Aside from the mimetic constraint provided in Eq.~(\ref{eq:mimeticconstraint}), which can be obtained by extremizing Eq.~(\ref{eq:action}) with respect to the Lagrange multiplier $\lambda$, the other field equations are obtained by extremizing Eq.~(\ref{eq:action}) with respect to the mimetic field $\phi$ and the metric tensor $g_{\mu\nu}$. This procedure leads to the following equation of motion for the scalar field:
\begin{eqnarray}
\partial_{\mu} \left ( \sqrt{-g}\lambda g^{\mu\nu}\partial_{\nu}\phi \right ) =0\,,
\label{eq:scalar}
\end{eqnarray}
and the following gravitational field equations:
\begin{eqnarray}
R_{\mu\nu}-\frac{1}{2}g_{\mu\nu}R=-2\lambda\partial_{\mu}\phi\partial_{\nu}\phi\,,
\label{eq:gravity}
\end{eqnarray}
with $R_{\mu\nu}$ and $R$ the Ricci tensor and Ricci scalar respectively. It is clear that Eq.~(\ref{eq:gravity}) differs from the field equations of GR through the extra source term on the right-hand side, which on cosmological scales mimics a cold DM component whose energy density is controlled by $\lambda$~\cite{Sebastiani:2016ras}.

It is important to note that the principle of conformal invariance allows us to fix the functional form of the conformal factor in terms of the auxiliary metric ${\tilde g}_{\mu\nu}$ and the scalar field $\phi$, but not the overall sign~\cite{Deruelle:2014zza,Gorji:2020ten}): therefore, one is in principle faced with two equally valid actions within the basic theory of mimetic gravity. For cosmological applications, in the presence of time-dependent homogeneous backgrounds, one typically adopts the $S_-$ action, where the mimetic field plays the role of dust (cold DM). However, consistent static solutions such as those describing compact objects require one to adopt the $S_+$ action, in order to avoid an imaginary mimetic field. Finally, relatively simple extensions of Eq.~(\ref{eq:action}) have been studied over the past decade, and allow one to mimic both DM and DE on cosmological scales (potentially also with an inflationary mechanism), and more generally to obtain any desired background expansion~\cite{Chamseddine:2014vna,Chaichian:2014qba,Nojiri:2014zqa,Myrzakulov:2015qaa,Astashenok:2015haa,Rabochaya:2015haa,Vagnozzi:2017ilo,Gorji:2017cai,Kaczmarek:2021psy,Benisty:2021cin}, making mimetic gravity an appealing cosmological framework.

\begin{figure*}[!ht]
\centering
\includegraphics[width=0.43\linewidth]{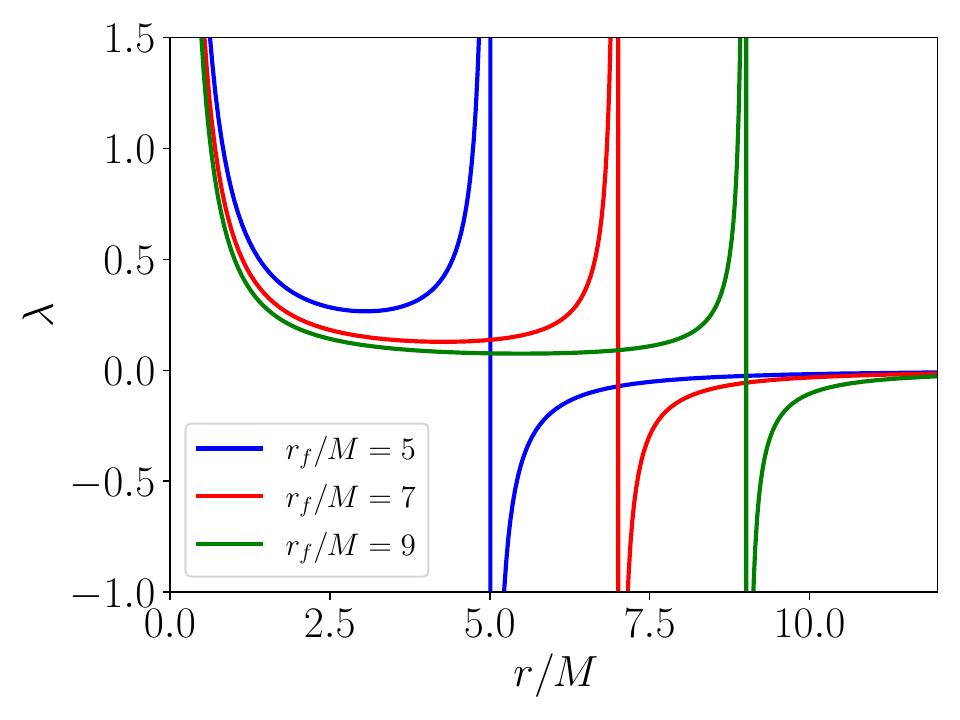}\,\,\,\,\,\includegraphics[width=0.43\linewidth]{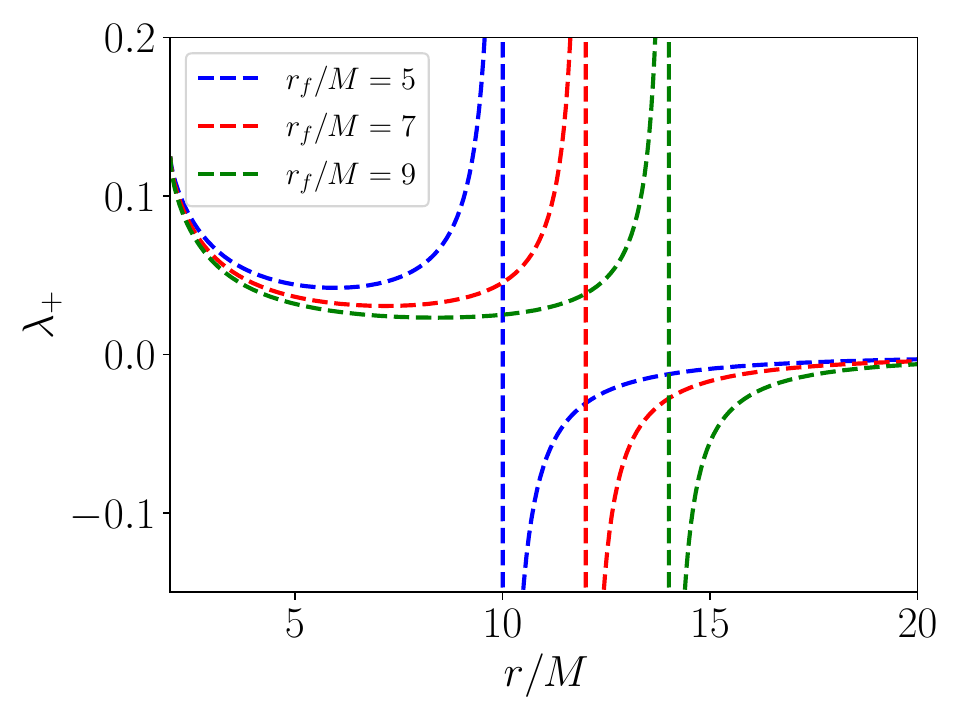}
\caption{\textbf{Radial profile of the Lagrange multiplier $\lambda$}. \textit{Left panel}: evolution of $\lambda$ as a function of the radial coordinate $r$ for the mimetic naked singularity, as given by Eq.~(\ref{eq:lambdans}), for different values of the parameter $r_f$, as indicated by the color coding. \textit{Right panel}: as in the left panel, but for the mimetic black hole, as given by Eq.~(\ref{eq:lambdabhext}), with the restriction $r \geq 2M$. In both cases the Lagrange multiplier diverges at the caustic singularity $r=r_f$, limiting the domain of validity of the solutions to $r<r_f$.}
\label{fig:lambda}
\end{figure*}

In what follows, we shall exclusively consider the case where $\lambda \neq 0$, since the $\lambda=0$ case is of no cosmological interest (in the latter case mimetic gravity admits stealth Schwarzschild BHs as well as related solutions~\cite{Oikonomou:2016fxb}, which are phenomenologically equivalent, including at the level of shadow properties, to Schwarzschild BHs). When attempting to construct static solutions in mimetic gravity, i.e.,\ where $\partial_t\phi=0$, an immediate obstruction emerges within the $S_-$ branch. In this case, it is clear that for the mimetic constraint Eq.~(\ref{eq:mimeticconstraint}) to be satisfied, the mimetic field must be imaginary. Although static spherically solutions in mimetic gravity with an imaginary field have been studied earlier~\cite{Myrzakulov:2015sea,Myrzakulov:2015kda,Sheykhi:2019gvk}, it was later realized that such solutions are pathological and cannot be considered astrophysically relevant. These results are completely consistent with a related no-go theorem in Ho\v{r}ava-Lifshitz gravity~\cite{Izumi:2009ry}, whose projectable version in the low-energy limit is equivalent to mimetic gravity~\cite{Cognola:2016gjy,Ramazanov:2016xhp,Chamseddine:2019gjh}. Therefore, the only way by means of which the static ansatz can be consistent with the mimetic constraint is by working in the $S_+$ branch and considering spacelike $\partial_{\mu}\phi$, with $\phi=\phi(r)$~\cite{Nashed:2018qag,Nashed:2021ctg,Nashed:2021hgn}. The impossibility of obtaining static solutions purely within the $S_-$ branch is consistent with the fact that in mimetic theories energy flows along $\partial_{\mu}\phi$ along this branch, which is a timelike geodesic~\cite{Lim:2010yk}. Given that energy freely falls, it is impossible to find static solutions.

With the above considerations in mind, the only static spherically symmetric solution supported within the $S_+$ branch of mimetic gravity when $\lambda \neq 0$ was shown to be a naked singularity, whose line element is the following~\cite{Gorji:2020ten}:
\begin{eqnarray}
ds_{\text{ns}}^2=-\frac{dt^2}{4 r^4 \lambda^2} + \frac{dr^2}{1+\frac{c_0}{r}}+r^2d\Omega^2 \,,
\label{eq:metricns}
\end{eqnarray}
where $d\Omega^2=d\theta^2 +\sin^2(\theta) d\varphi^2$ is the metric on the 2-sphere, $c_0>0$ is an integration constant, and the Lagrange multiplier $\lambda(r)$ takes the following form:
\begin{eqnarray}
\lambda(r) = \frac{1}{2r^2} \left [ 1 - \sqrt{1+\frac{c_0}{r}} \times \left ( \frac{1}{\sqrt{1+\frac{c_0}{r_f}}} +\ln\sqrt{\frac{r}{r_f}}\frac{1+\sqrt{1+\frac{c_0}{r}}}{1+\sqrt{1+\frac{c_0}{r_f}}}\,\right ) \, \right ] ^{-1}\,,
\label{eq:lambdans}
\end{eqnarray}
with $r_f$ being a free parameter. The behavior of $\lambda(r)$ is shown in the left panel of Fig.~\ref{fig:lambda} for different values of $r_f$, making it clear that $\lambda$ diverges at both $r=0$ and $r=r_f$: the former is associated to the usual central curvature singularity present also in the Schwarzschild BH, whereas the latter is associated to a caustic singularity~\cite{Gorji:2020ten}. Given that for $r>r_f$ the curvature invariants switch sign, we consider only the region of positive curvature and restrict the coordinate $r$ to $r \in [0, r_f)$, where the metric is applicable, and avoiding the caustic singularity~\cite{Gorji:2020ten}.

The NS solution described previously assumes the $S_+$ branch throughout the entire manifold, and requires $c_0>0$ in Eq.~(\ref{eq:metricns}). However, in the $c_0<0$ case it is possible to construct a mimetic BH solution~\cite{Gorji:2020ten}. This requires a particular gluing technique: the $S_+$ branch is considered when describing the exterior space-time, whereas the interior space-time described by the $S_-$ branch is homogeneous but anisotropic, with the two solutions glued through appropriate junction conditions. The line element of the resulting mimetic BH exterior space-time, in coordinates $(t_+,r_+,\theta,\varphi)$, is given by~\cite{Gorji:2020ten}:
\begin{eqnarray}
ds_{\text{bh}+}^2=-\frac{dt_+^2}{4r_+^4\lambda_+^2}+\frac{dr_+^2}{1-\frac{2m}{r_+}}+r_+^2 d\Omega^2 \,,
\label{eq:metricbhext}
\end{eqnarray}
where $m_+=-c_0/2$ can ultimately be identified with the BH mass $M$ [so Eq.~(\ref{eq:metricbhext}) is effectively of the same form of Eq.~(\ref{eq:metricns})], whereas the Lagrange multiplier $\lambda_+$ takes the same form as Eq.~(\ref{eq:lambdans}):
\begin{eqnarray}
\lambda_+(r_+) = \frac{1}{2r_+^2} \left [ 1 - \sqrt{1-\frac{2m_+}{r_+}} \left ( \frac{1}{\sqrt{1-\frac{2m_+}{r_f}}} +\ln\sqrt{\frac{r_+}{r_f}}\frac{1+\sqrt{1-\frac{2m_+}{r_+}}}{1+\sqrt{1-\frac{2m_+}{r_f}}}\,\right ) \, \right ] ^{-1}\,,
\label{eq:lambdabhext}
\end{eqnarray}
with $r_f$ being once more a free parameter. This solution is valid for $r_+ \geq \vert c_0 \vert = 2m_+$, and the behaviour of $\lambda_+(r)$ is shown in the right panel of Fig.~\ref{fig:lambda} for different values of $r_f$. The behaviour of $\lambda_+(r)$ is obviously identical to that of $\lambda(r)$ for the NS, with the only key difference being the restricted domain in the $r$ coordinate. In this way, the central singularity is avoided, but the caustic singularity at $r=r_f$ remains~\cite{Gorji:2020ten}. The metric is therefore applicable in the range $r \in [2m_+, r_f)$, and the surface $r_+=2m_+$ constitutes an apparent horizon for the mimetic BH~\cite{Gorji:2020ten}.

For completeness, although it is not required for the purposes of studying the mimetic BH's shadow properties, we also report the line element of the mimetic BH interior space-time, matched continuously to that in Eq.~(\ref{eq:lambdabhext}) at the apparent horizon. In coordinates $(t_-,r_-,\theta,\varphi)$, this time-dependent homogeneous anisotropic line element is given by~\cite{Gorji:2020ten}:
\begin{eqnarray}
ds_{\text{bh}-}^2=-\frac{dt_-^2}{\frac{2m_-}{t}-1} + \frac{dr_-^2}{4t_-^4\lambda_-^2}+t_-^2d\Omega^2\,,
\label{eq:metricbhint}
\end{eqnarray}
where $m_-=m_+=-2c_0$ and will be denoted by $M$ in what follows, whereas the Lagrange multiplier $\lambda_-$ is given by:
\begin{eqnarray}
\lambda_-(r) = -\frac{1}{2t_-^2} \left \{ 1 - \sqrt{\frac{2m_-}{t_-}-1} \left [ \frac{1}{\sqrt{\frac{2m_-}{t_i}}-1}+\arctan \left ( \sqrt{\frac{2m_-}{t_-}-1} \right ) -\arctan \left ( \sqrt{\frac{2m_-}{t_i}-1} \right ) \right ] \right \} ^{-1}\,,
\label{eq:lambdabhint}
\end{eqnarray}
where $t_i$ is one of the two times at which $\lambda_-$ diverges (the other one being at $t_-=0$). The complete mimetic BH solution is obtained by gluing Eq.~(\ref{eq:metricbhext}) and Eq.~(\ref{eq:metricbhint}) at the apparent horizon. Nevertheless, as mentioned earlier, it is only the exterior solution which is relevant for studying the mimetic BH shadow properties, whereas the interior solution has only been provided for completeness.

\section*{Methods: shadow properties of mimetic compact objects}
\label{sec:shadow}

The optical appearance of SMBHs in Very Long Baseline Interferometry (VLBI) arrays such as the EHT is closely tied to the theoretical concept of BH shadow~\cite{Cunha:2018acu,Perlick:2021aok,Chen:2022scf}. The main features observed in VLBI BH images are a bright emission ring surrounding a central brightness depression~\cite{Falcke:1999pj}, and it is the latter which is related to the BH shadow. The edge of the BH shadow, on the plane of a distant observer, separates capture orbits from scattering orbits, and is the apparent image of the photon region (photon sphere in the case of spherically symmetric metrics), i.e.\ the region of space-time which supports closed photon orbits. In recent years, following the milestone horizon-scale images of M87$^{\star}$~\cite{EventHorizonTelescope:2019dse} and Sgr A$^{\star}$~\cite{EventHorizonTelescope:2022wkp} released by the EHT, there has been enormous interest in using BH shadows to test fundamental physics~\cite{Held:2019xde,Vagnozzi:2019apd,Zhu:2019ura,Cunha:2019ikd,Banerjee:2019nnj,Banerjee:2019xds,Zhdanov:2019ozq,Allahyari:2019jqz,Khodadi:2020jij,Kumar:2020yem,Khodadi:2020gns,Pantig:2021zqe,EventHorizonTelescope:2021dqv,Khodadi:2021gbc,Stashko:2021lad,Uniyal:2022vdu,Pantig:2022ely,Ghosh:2022kit,Khodadi:2022pqh,KumarWalia:2022aop,Shaikh:2022ivr,Afrin:2022ztr,Pantig:2023yer,Gonzalez:2023rsd,Sahoo:2023czj,Nozari:2023flq,Uniyal:2023ahv,Filho:2023ycx,EventHorizonTelescope:2022xqj,Raza:2023vkn,Hoshimov:2023tlz,Chakhchi:2024tzo,Liu:2024lve,Liu:2024lbi}.

The line element of the space-times we are interested in are all of the following form:
\begin{eqnarray}
ds^2=-A(r)dt^2+B(r)dr^2+r^2d\Omega^2 \,,
\label{eq:metricgeneral}
\end{eqnarray}
where we refer to $A(r)$ as the lapse function (also referred to as metric function in various works in the shadow literature), and $r$ is manifestly the areal radius. If such a metric admits a photon sphere, its radial coordinate $r_{\text{ph}}$ is given by the solution to~\cite{Perlick:2021aok}:
\begin{eqnarray}
A(r_{\text{ph}})-\frac{1}{2}r_{\text{ph}}A'(r_{\text{ph}})=0\,,
\label{eq:rph}
\end{eqnarray}
where the prime denotes a derivative with respect to $r$. The radius of the shadow seen by a distant observer, $r_{\text{sh}}$, is given by:
\begin{eqnarray}
r_{\text{sh}}=\frac{r}{\sqrt{A(r)}}\Bigg\vert_{r_{\text{ph}}}\,.
\label{eq:rsh}
\end{eqnarray}
In what follows, we will use Eqs.~(\ref{eq:rph},\ref{eq:rsh}) to compute the size of the shadows cast by mimetic compact objects, reading off $A(r)$ from Eqs.~(\ref{eq:metricns},\ref{eq:metricbhext}) for the mimetic NS and mimetic BH respectively.

\subsection*{Mimetic naked singularity}
\label{subsec:mimeticnakedsingularity}

\begin{figure*}[!t]
\centering
\includegraphics[width=0.33\linewidth]{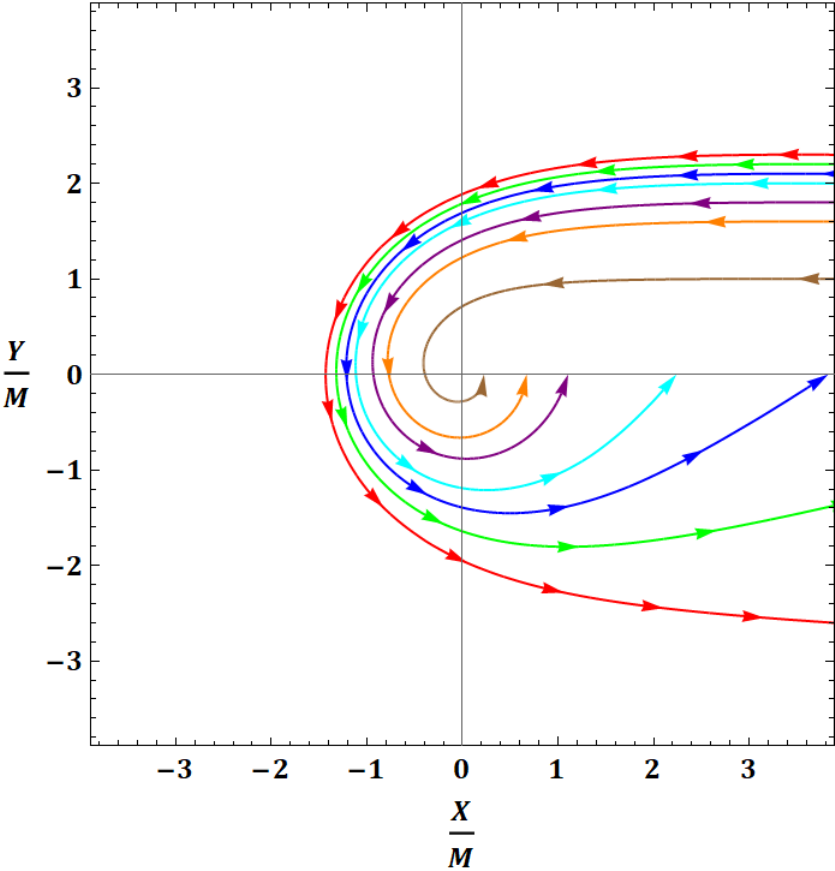}\,\,\,\,\,\includegraphics[width=0.33\linewidth]{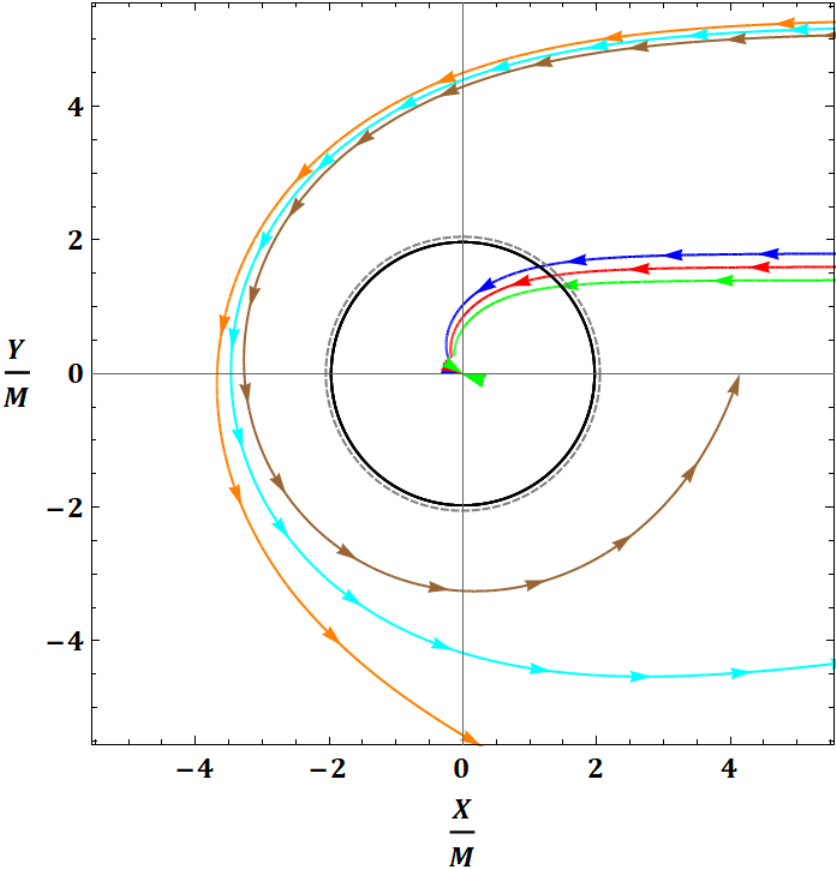}\,\,\,\,\,\includegraphics[width=0.33\linewidth]{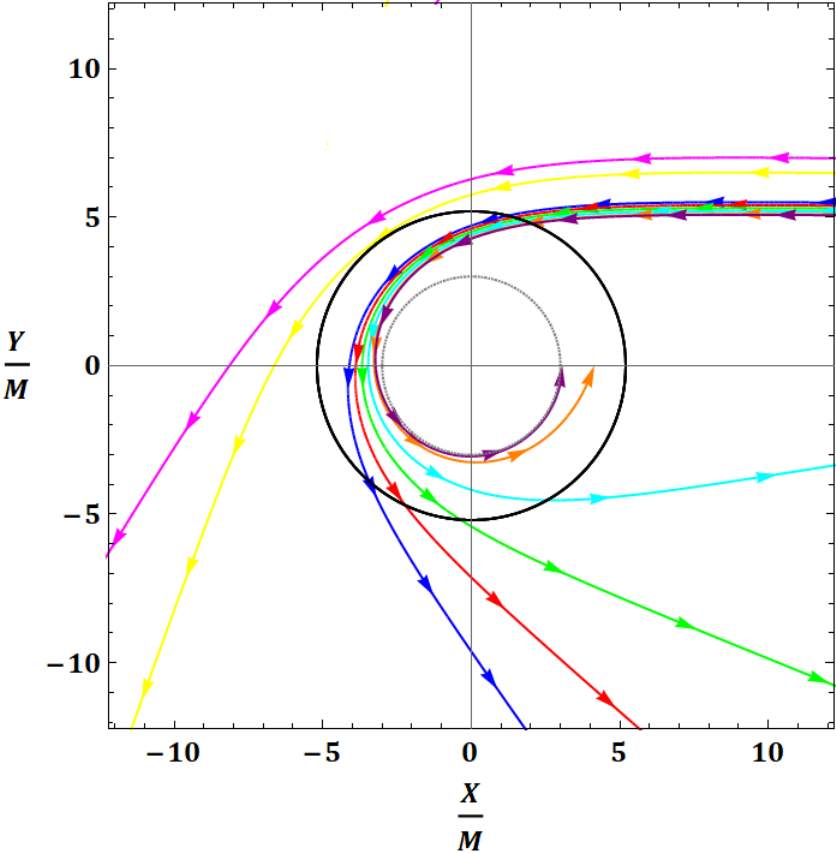}
\caption{\textbf{Null geodesics}. Examples of null geodesics in the equatorial plane of the mimetic naked singularity (\textit{left panel}), mimetic black hole (\textit{intermediate panel}), and Schwarzschild black hole (\textit{right panel}). Within each panel, colors correspond to null geodesics characterized by different impact parameters. The black solid and grey dotted curves indicate the shadow and photon sphere respectively (as discussed in the text, both are absent in the case of the mimetic naked singularity, for which only scattering orbits exist).}
\label{fig:null}
\end{figure*}

We begin by considering the mimetic NS, for which $A(r)=1/4r^4\lambda^2$, and for which we recall that the domain of validity is $r \in [0,r_f)$. Using the fact that $\lambda(r)$ is given by Eq.~(\ref{eq:lambdans}), and plugging $A(r)=1/4r^4\lambda(r)^2$ into Eq.~(\ref{eq:rph}), with some tedious but otherwise completely straightforward algebraic manipulation Eq.~(\ref{eq:rph}) can be recast into the following form:
\begin{eqnarray}
f_1^2(r_{\text{ph}})-r_{\text{ph}}f_1(r_{\text{ph}})f_2(r_{\text{ph}})=0\,,
\label{eq:rphns}
\end{eqnarray}
with the functions $f_1(r)$ and $f_2(r)$ given by:
\begin{eqnarray}
f_1(r_{\text{ph}})=1-\sqrt{1+\frac{c_0}{r_{\text{ph}}}} \left [ \frac{1}{\sqrt{1+\frac{c_0}{r_f}}}+\ln \left ( \frac{ \left ( 1+\sqrt{1+\frac{c_0}{r_{\text{ph}}}} \right ) \sqrt{\frac{r_{\text{ph}}}{r_f}}}{1+\sqrt{1+\frac{c_0}{r_f}}} \right ) \right ]^2 \,,
\label{eq:f1}
\end{eqnarray}
and:
\begin{equation}
\begin{split}
&f_2(r_{\text{ph}})=\frac{c_0 \left \{ \frac{1}{\sqrt{1+\frac{c_0}{r_f}}}+\ln \left [ \frac{ \left ( 1+\sqrt{1+\frac{c_0}{r_{\text{ph}}}} \right ) \sqrt{\frac{r_{\text{ph}}}{r_f}}}{1+\sqrt{1+\frac{c_0}{r_f}}} \right ] \right \} }{2 r_{\text{ph}}^2 \sqrt{1+\frac{c_0}{r_{\text{ph}}}}}- \left [ \left ( 1+\sqrt{1+\frac{c_0}{r_{\text{ph}}}} \right ) \sqrt{\frac{r_{\text{ph}}}{r_f}} \right ] ^{-1} \\
&\times \left [ \sqrt{1+\frac{c_0}{r_{\text{ph}}}} \left ( 1+\sqrt{1+\frac{c_0}{r_f}} \right ) \right ] \left [ \frac{1+\sqrt{1+\frac{c_0}{r_{\text{ph}}}}}{2r_f \left ( 1+\sqrt{1+\frac{c_0}{r_f}} \right ) \sqrt{\frac{r_{\text{ph}}}{r_f}}}- \frac{c_0\sqrt{\frac{r_{\text{ph}}}{r_f}}}{2r_{\text{ph}}^2 \sqrt{1+\frac{c_0}{r_{\text{ph}}}} \left ( 1+\sqrt{1+\frac{c_0}{r_f}} \right ) } \right ] \,.
\end{split}
\label{eq:f2}
\end{equation}
We solve Eq.~(\ref{eq:rphns}) numerically for various values of $c_0$ and $r_f$. We find that, regardless of the value of $c_0>0$, there are always two roots for the photon sphere radius $r_{\text{ph}}$: one exactly at $r_{\text{ph},1}=r_f$, and the second at $r_{\text{ph},2}>r_f$. The value of the first root is always $r_f$ independently of the value of $c_0$, whereas the second varies as $c_0$ changes, but always remains $>r_f$. We recall that for the mimetic NS the domain of validity is $r \in [0,r_f)$, which implies that neither of the two roots are physical. Therefore, the mimetic NS does not possess a photon sphere in the domain of validity.

One might be tempted to jump to the conclusion that the mimetic NS does not cast a shadow, given the lack of photon sphere. However, explicit counterexamples of NSs which do not possess a photon sphere, and yet cast a shadow, have been constructed in the literature~\cite{Joshi:2020tlq,Dey:2020bgo}. A space-time with no photon sphere may still cast a shadow, though whether or not it does is critically related to the behaviour of the effective potential $V_{\text{eff}}(r)$ near the origin, and in particular on whether the specific form of $V_{\text{eff}}(r)$ supports a minimum impact parameter of turning point for null geodesics which is non-zero. This is possible if and only if the effective potential does not diverge. Denoting by $b_{\text{tp}}$ and $r_{\text{tp}}$ the minimum impact parameter and turning point coordinate for null geodesics, the following holds for static spherically symmetric metrics:
\begin{eqnarray}
V_{\text{eff}}(r_{\text{tp}})=1/b_{\text{tp}}^2=1/r_{\text{sh}}^2\,,
\label{eq:turning}
\end{eqnarray}
which clearly shows that if $V$ diverges, $b_{\text{tp}}\rightarrow 0$ and $r_{\text{sh}} \rightarrow 0$. To put it differently, if $b_{\text{tp}} \neq 0$, null geodesics coming from a distant source with impact parameter $b<b_{\text{tp}}$ will end up being trapped closer to the singularity, whereas those with $b>b_{\text{tp}}$ will manage to scatter back to infinity: this leads to a shadow feature being seen by a distant observer even in the absence of a photon sphere, solely as a result of the curvature of space-time~\cite{Joshi:2020tlq}. In the case of the mimetic NS, it is easy to see that the effective potential diverges, so the impact parameter at the minimum turning point radius, $r_{\text{tp}}=0$, is given by~\cite{Joshi:2020tlq}:
\begin{eqnarray}
b_{\text{tp}} = \frac{r}{\sqrt{A(r)}}\Bigg\vert_{r_{\text{tp}}} \rightarrow 0\,.
\label{eq:btp}
\end{eqnarray}
This is in contrast to other examples studied in the literature where, despite the absence of a photon sphere, $b_{\text{tp}}$ converges to a finite, non-zero value~\cite{Joshi:2020tlq,Vagnozzi:2022moj}. We therefore conclude that \textit{the mimetic naked singularity does not cast a shadow}. From this, we can immediately assert that mimetic NSs are ruled out as candidates to explain the EHT images of M87$^{\star}$ and Sgr A$^{\star}$. This can also be appreciated in the left panel of  Fig.~\ref{fig:null}, where we show examples of null geodesics in the equatorial plane of the mimetic NS. It is clear that the mimetic NS only possesses scattering orbits, even for low impact parameter values, because the divergent effective potential ``repels'' null geodesics away from the singularity. This, in turn, leads to the mimetic NS not casting a shadow. For comparison, in the intermediate panel of Fig.~\ref{fig:null} we show examples of null geodesics in the equatorial plane of the mimetic BH (to be discussed shortly), whereas in the right panel we do the same for the Schwarzschild BH: Fig.~\ref{fig:null} clearly shows the very different nature of null geodesics in all three space-times, which in turn leads to their having very different shadow properties.

Part of the above result was predicated upon the absence of a photon sphere, and in particular our choice to discard the first root $r_{\text{ph},1}=r_f$ on the basis of the domain of validity being $r \in [0,r_f)$. One may however wonder if and how the above results change if we admit $r_{\text{ph}}=r_f$ as a legitimate root. Recall that $\lambda$ diverges at the caustic singularity $r=r_f$ (see the left panel of Fig.~\ref{fig:lambda}), and therefore $A(r)=1/4r^4\lambda^2 \to 0$, implying that $r_{\text{sh}}$ would diverge. This is of course phenomenologically unacceptable and in stark disagreement with the EHT images of M87$^{\star}$ and Sgr A$^{\star}$ (although we stress that this result is anyhow inadmissible as $r_{\text{ph},1}=r_f$ is not a valid root).

\subsection*{Mimetic black hole}
\label{subsec:mimeticblackhole}

\begin{figure*}[!t]
\centering
\includegraphics[width=0.43\linewidth]{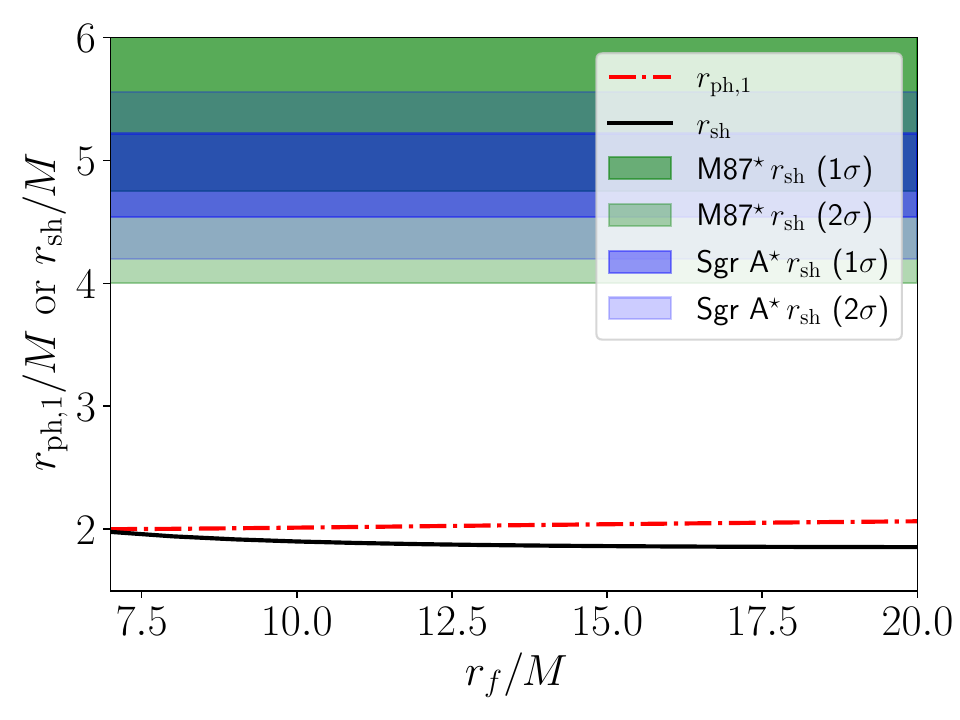}\,\,\,\,\,\includegraphics[width=0.43\linewidth]{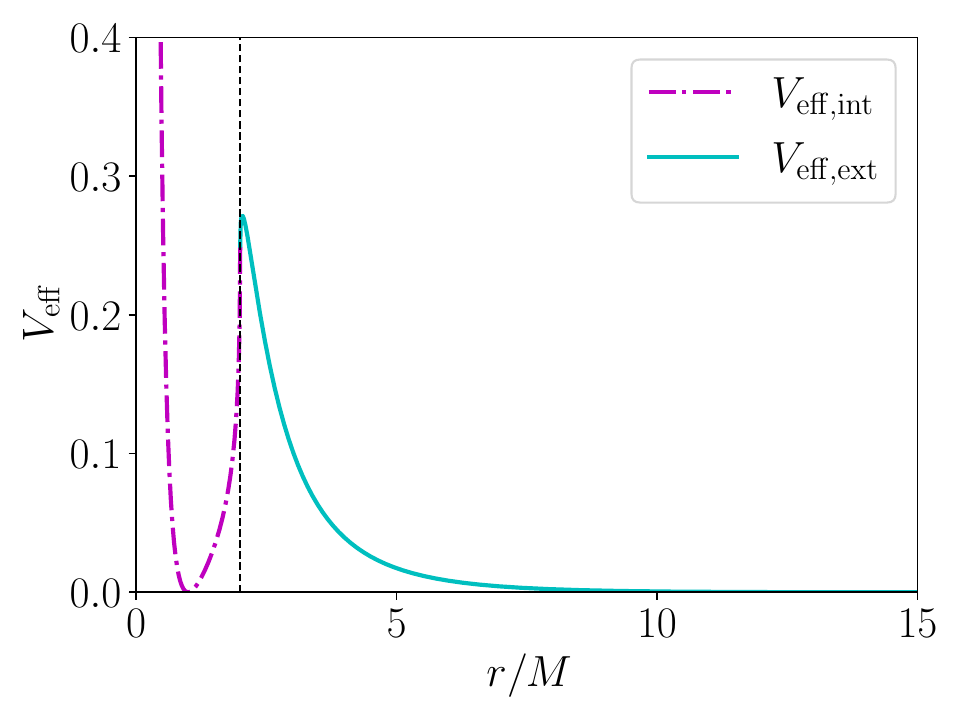}
\caption{\textbf{Mimetic black hole photon sphere, shadow, and effective potential}. \textit{Left panel}: evolution of the size of the photon ring (red dot-dashed curve) and shadow (black solid curve) of the mimetic black hole as a function of the parameter $r_f$. The dark green and light green (dark blue and light blue) regions are consistent with the EHT image of M87$^{\star}$ [Sgr A$^{\star}$] at $1\sigma$ and $2\sigma$ respectively. \textit{Right panel}: Effective potential profile of the mimetic black hole space-time, for both the exterior (cyan solid curve) and interior (magenta dot-dashed curve) solutions, with the dashed vertical curve indicating the location of the apparent horizon at $r_H=2M$.}
\label{fig:rph1rsheffectivepotential}
\end{figure*}

We now move on to the mimetic BH. To study its shadow properties we only need to consider the exterior space-time, for which $A(r)=1/4r^4\lambda_+^2$, and for which we recall that the domain of validity is $r \in [2M,r_f)$. Using the fact that $\lambda_+(r)$ is given by Eq.~(\ref{eq:lambdabhext}), and plugging $A(r)=1/4r^4\lambda_+(r)^2$ into Eq.~(\ref{eq:rph}), with some tedious but otherwise completely straightforward algebraic manipulation Eq.~(\ref{eq:rph}) can be recast into the following form:
\begin{eqnarray}
g_1^2(r_{\text{ph}})-r_{\text{ph}}g_1(r_{\text{ph}})g_2(r_{\text{ph}})=0\,,
\label{eq:rphbh}
\end{eqnarray}
with the functions $g_1(r)$ and $g_2(r)$ given by:
\begin{eqnarray}
g_1(r_{\text{ph}})=1-\sqrt{1-\frac{2m}{r_{\text{ph}}}} \left [ \frac{1}{\sqrt{1-\frac{2m}{r_f}}}+\ln \left ( \frac{ \left ( 1+\sqrt{1-\frac{2m}{r_{\text{ph}}}} \right ) \sqrt{\frac{r_{\text{ph}}}{r_f}}}{1+\sqrt{1-\frac{2m}{r_f}}}\right) \right ] ^2\,,
\label{eq:g1}
\end{eqnarray}
and:
\begin{equation}
\begin{split}
&g_2(r_{\text{ph}})=-\frac{m \left \{ \frac{1}{\sqrt{1-\frac{2m}{r_f}}}+\ln \left [ \frac{ \left ( 1+\sqrt{1-\frac{2m}{r_{\text{ph}}}} \right ) \sqrt{\frac{r_{\text{ph}}}{r_f}}}{1+\sqrt{1-\frac{2m}{r_f}}} \right ] \right \} }{2 r_{\text{ph}}^2 \sqrt{1-\frac{2m}{r_{\text{ph}}}}}- \left [ \left ( 1+\sqrt{1-\frac{2m}{r_{\text{ph}}}} \right ) \sqrt{\frac{r_{\text{ph}}}{r_f}} \right ] ^{-1} \\
&\times \left [ \sqrt{1-\frac{2m}{r_{\text{ph}}}} \left ( 1+\sqrt{1-\frac{2m}{r_f}} \right ) \right ] \left [ \frac{1+\sqrt{1+\frac{2m}{r_{\text{ph}}}}}{2 r_f \left ( 1+\sqrt{1-\frac{2m}{r_f}} \right ) \sqrt{\frac{r_{\text{ph}}}{r_f}}}+\frac{m\sqrt{\frac{r_{\text{ph}}}{r_f}}}{2 r_{\text{ph}}^2 \sqrt{1-\frac{2m}{r_{\text{ph}}}} \left ( 1+\sqrt{1-\frac{2m}{r_f}}\right ) } \right ] \,.
\end{split}
\label{eq:g2}
\end{equation}
We solve Eq.~(\ref{eq:rphns}) numerically for various values of $r_f$.

From our numerical analysis, we observe that the behavior of the roots of Eq.~(\ref{eq:rphns}) depends on whether $r_f \gtrless 7M$. For $r_f \leq 7M$ we find a unique root at $r_{\text{ph}}=r_f$. This is of course physically unacceptable as it falls out of the domain of validity $r \in [2M,r_f)$ (and, just as in the NS case, accepting it would lead to a divergent $r_{\text{sh}}$ which is again phenomenologically unacceptable). On the other hand, for $r_f>7M$, we find two distinct roots. The larger root is always located at $r_{\text{ph},2}=r_f$ and, as per our earlier considerations, has to be discarded. On the other hand, the smaller root $r_{\text{ph},1}$ lies in the range $r \in [2M,r_f)$ and hence can be considered physically valid.

The location of the physically valid root $r_{\text{ph},1}$ as a function of $r_f$ for $r_f>7M$ is given by the red dot-dashed curve in the left panel of Fig.~\ref{fig:rph1rsheffectivepotential}. The trend we observe is that $r_{\text{ph},1}$ is very close to the radius of the apparent horizon, $r_H=2M$, and is a monotonically (extremely slowly) increasing function of $r_f$. To better understand the origin of this root, we plot an example effective potential $V_{\text{eff}}(r)$ in the right panel of Fig.~\ref{fig:rph1rsheffectivepotential}. With no loss of generality and consistently with the symmetries of the problem, we consider motion in the equatorial plane, in which case $V_{\text{eff}}=A(r)/r^2$. We see that as $r$ decreases moving towards the horizon, $V_{\text{eff}}$ increases sharply, before displaying a small dip close to the horizon. Being a local maximum, the peak in $V_{\text{eff}}$ marks the position of unstable circular orbits, and hence of the photon sphere, which lies very close to the horizon. This can be contrasted to the typical shape of the effective potential for the Schwarzschild BH, where the location of the photon sphere $r_{\text{ph}}=3M$ is well distinct from that of the event horizon.

Taking the physically acceptable root of Eq.~(\ref{eq:rphbh}), we use Eq.~(\ref{eq:rsh}) to calculate the size of the resulting shadow, given by the black solid curve in the left panel of Fig.~\ref{fig:rph1rsheffectivepotential}. Our calculation shows that $r_{\text{sh}}$ is a monotonically (very slowly) decreasing function of $r_f$, decreasing from values $r_{\text{sh}} \approx 2M$ for $r_f \approx 7M$, to progressively smaller values as $r_f$ is increased (see also the intermediate panel of Fig.~\ref{fig:null}). Again, this can be contrasted to the case of the Schwarzschild BH, where $r_{\text{sh}}=3\sqrt{3}M$ (see the right panel of Fig.~\ref{fig:null}).

These values of $r_{\text{sh}}$ are observationally problematic as they correspond to a shadow size much smaller than what has been inferred for both M87$^{\star}$ and Sgr A$^{\star}$, both of which are consistent with expectations for a Schwarzschild BH given external measurements of the sources' mass-to-distance ratios. For instance, in the case of M87$^{\star}$, combining information on the angular size of the image and the distance to the source, a shadow diameter $2r_{\text{sh}}/M \approx 11.0 \pm 1.5$ is found~\cite{Bambi:2019tjh} (see the blue bands in Fig.~\ref{fig:rph1rsheffectivepotential}), in agreement with the expected $6\sqrt{3} \approx 10.4$ for the Schwarzschild metric. Similarly, for Sgr A$^{\star}$, combining measurements of the source's mass-to-distance ratio from Keck and VLTI, the $2\sigma$ interval $4.21 \lesssim r_{\text{sh}}/M \lesssim 5.56$ is inferred~\cite{Vagnozzi:2022moj} (see the green bands in Fig.~\ref{fig:rph1rsheffectivepotential}). In both cases, the predicted size of the mimetic BH shadow is outside of the allowed range by several standard deviations, as is clear from the blue and green bands in Fig.~\ref{fig:rph1rsheffectivepotential}. We therefore conclude that, despite mimetic BHs casting a shadow, the latter is too small to be consistent with the images of M87$^{\star}$ and Sgr A$^{\star}$ observed by the EHT, which therefore exclude these space-times in the domain of validity. We conclude that black hole shadow observations rule out baseline mimetic gravity.

\section*{Conclusions}
\label{sec:conclusions}

\begin{figure*}[!t]
\centering
\includegraphics[width=0.75\linewidth]{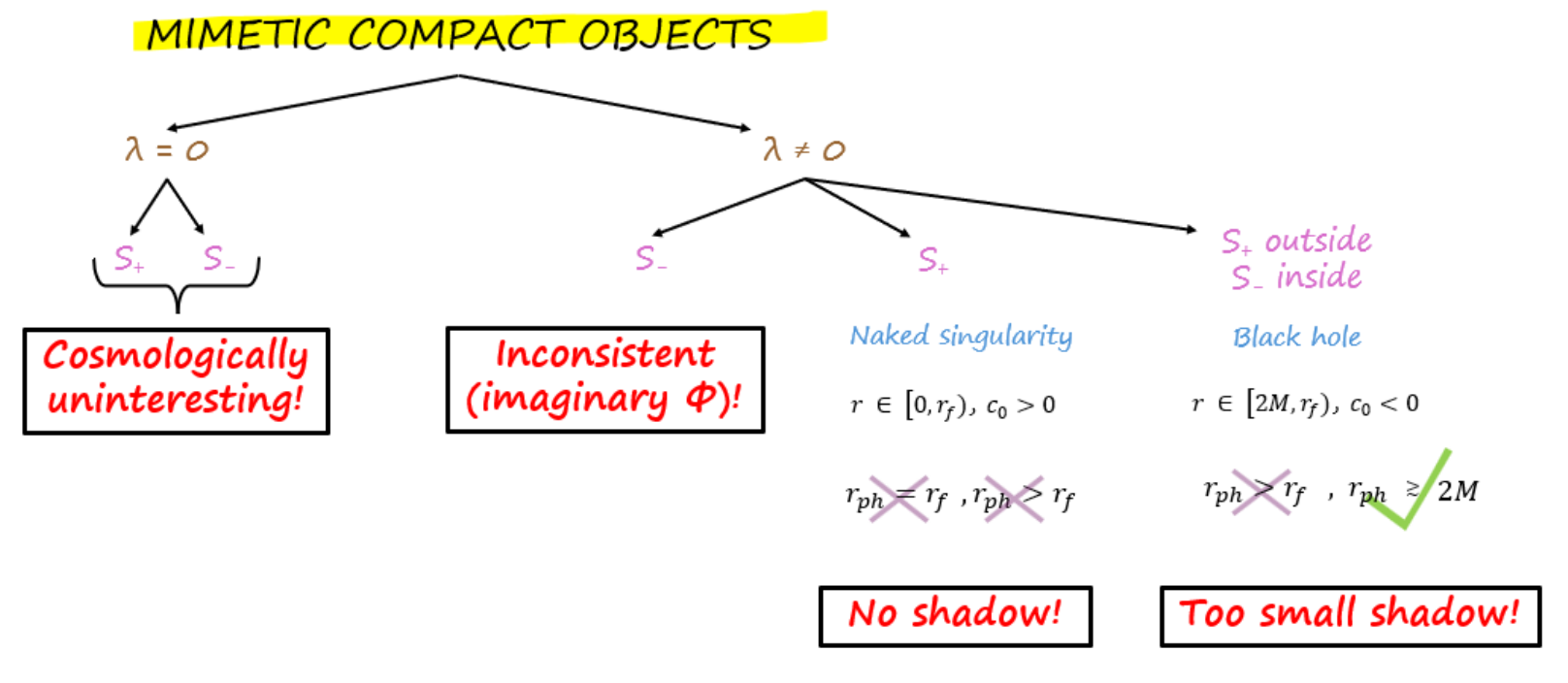}
\caption{\textbf{Summary figure}. Diagrammatic synopsis of the main results of this paper.}
\label{fig:diagrammaticsummary}
\end{figure*}

A number of theories of modified gravity are cosmologically very appealing in light of their potential to account, albeit not without challenges, for the dark sector components. It is nonetheless of paramount importance to test the consistency of these models against the increasing range of available observations on smaller scales. We considered the case of mimetic gravity~\cite{Chamseddine:2013kea,Sebastiani:2016ras}, a framework of modified gravity related to GR by a non-invertible disformal transformation, and which can mimic dark matter and dark energy on cosmological scales. Consistent solutions for compact objects in the baseline mimetic gravity setting are non-trivial, and the only natural objects in this sense are a naked singularity solution, and a black hole obtained through a non-trivial gluing procedure. Our goal in this work has been to study the shadow properties of these non-trivial objects, in order to qualitatively confront them against recent VLBI horizon-scale images from the Event Horizon Telescope.

We have shown that the shadow properties of both classes of compact objects in mimetic gravity are pathological: in particular, the mimetic naked singularity does not cast a shadow, whereas the mimetic BH always casts a shadow which is significantly smaller than that of the Schwarzschild space-time. A diagrammatic summary of our results is provided in Fig.~\ref{fig:diagrammaticsummary}. Even without going in detail into an analysis of the optical appearance of these compact objects (which would require a model for the surrounding accretion flow), the above properties are more than sufficient to conclude that both space-times are in very strong disagreement with the EHT images of M87$^{\star}$ and Sgr A$^{\star}$~\cite{EventHorizonTelescope:2019dse,EventHorizonTelescope:2022wkp}.

In short, our main conclusion is that black hole shadow observations rule out baseline mimetic gravity. There remains the possibility of stealth Schwarzschild BHs in mimetic gravity, which would be in agreement with the EHT observations -- however, such a trivial solution is only allowed when $\lambda=0$, which is obviously cosmologically uninteresting. Another possibility is to envisage extensions and/or variations of baseline mimetic gravity, which may accommodate different classes of space-times (an interesting possibility in this sense is to modify the mimetic constraint by introducing a function $\omega(\phi)$ on the left-hand side~\cite{Nojiri:2022cah,Nojiri:2024txy}), and we leave a detailed investigation thereof to future work. Our results highlight an interesting complementarity between cosmology and BH imaging, where the latter observations can be used to rule out theories of interest to the former. At the dawn of the BH imaging era and as a number of Stage IV cosmological surveys have just launched, we strongly encourage the community to explore similar tests in order to narrow down viable theories of gravity.

\section*{Acknowledgements}

M.Kh. and S.V. acknowledge useful discussions with Sergei Odintsov and Sergio Zerbini. S.V. acknowledges support from the University of Trento through the UniTrento Internal Call for Research 2023 grant ``Searching for Dark Energy off the beaten track'' (DARKTRACK, grant agreement no.\ E63C22000500003), and from INFN through the CSN4 Iniziativa Specifica ``Quantum Fields in Gravity, Cosmology and Black Holes'' (FLAG).

\section*{Additional information}

The authors declare no competing interests.

\section*{Data availability}

The datasets used and/or analysed during the current study are available from the corresponding author on reasonable request.

\section*{Author contributions statement}

M.Kh. and S.V. conceived the project, M.Kh. carried out the relevant calculations, M.Kh. and S.V. analyzed the results, M.Kh. and S.V. produced the plots, M.Kh. and S.V. wrote the initial draft. All authors reviewed and contributed to writing the manuscript.

\bibliography{Mimetic}

\begin{thebibliography}{100}
\urlstyle{rm}
\expandafter\ifx\csname url\endcsname\relax
  \def\url#1{\texttt{#1}}\fi
\expandafter\ifx\csname urlprefix\endcsname\relax\def\urlprefix{URL }\fi
\expandafter\ifx\csname doiprefix\endcsname\relax\def\doiprefix{DOI: }\fi
\providecommand{\bibinfo}[2]{#2}
\providecommand{\eprint}[2][]{\url{#2}}

\bibitem{Nojiri:2010wj}
\bibinfo{author}{Nojiri, S.} \& \bibinfo{author}{Odintsov, S.~D.}
\newblock \bibinfo{journal}{\bibinfo{title}{{Unified cosmic history in modified
  gravity: from F(R) theory to Lorentz non-invariant models}}}.
\newblock {\emph{\JournalTitle{Phys. Rept.}}} \textbf{\bibinfo{volume}{505}},
  \bibinfo{pages}{59--144}, \doiprefix\url{10.1016/j.physrep.2011.04.001}
  (\bibinfo{year}{2011}).
\newblock \eprint{1011.0544}.

\bibitem{Clifton:2011jh}
\bibinfo{author}{Clifton, T.}, \bibinfo{author}{Ferreira, P.~G.},
  \bibinfo{author}{Padilla, A.} \& \bibinfo{author}{Skordis, C.}
\newblock \bibinfo{journal}{\bibinfo{title}{{Modified Gravity and Cosmology}}}.
\newblock {\emph{\JournalTitle{Phys. Rept.}}} \textbf{\bibinfo{volume}{513}},
  \bibinfo{pages}{1--189}, \doiprefix\url{10.1016/j.physrep.2012.01.001}
  (\bibinfo{year}{2012}).
\newblock \eprint{1106.2476}.

\bibitem{Nojiri:2017ncd}
\bibinfo{author}{Nojiri, S.}, \bibinfo{author}{Odintsov, S.~D.} \&
  \bibinfo{author}{Oikonomou, V.~K.}
\newblock \bibinfo{journal}{\bibinfo{title}{{Modified Gravity Theories on a
  Nutshell: Inflation, Bounce and Late-time Evolution}}}.
\newblock {\emph{\JournalTitle{Phys. Rept.}}} \textbf{\bibinfo{volume}{692}},
  \bibinfo{pages}{1--104}, \doiprefix\url{10.1016/j.physrep.2017.06.001}
  (\bibinfo{year}{2017}).
\newblock \eprint{1705.11098}.

\bibitem{Chamseddine:2013kea}
\bibinfo{author}{Chamseddine, A.~H.} \& \bibinfo{author}{Mukhanov, V.}
\newblock \bibinfo{journal}{\bibinfo{title}{{Mimetic Dark Matter}}}.
\newblock {\emph{\JournalTitle{JHEP}}} \textbf{\bibinfo{volume}{11}},
  \bibinfo{pages}{135}, \doiprefix\url{10.1007/JHEP11(2013)135}
  (\bibinfo{year}{2013}).
\newblock \eprint{1308.5410}.

\bibitem{Sebastiani:2016ras}
\bibinfo{author}{Sebastiani, L.}, \bibinfo{author}{Vagnozzi, S.} \&
  \bibinfo{author}{Myrzakulov, R.}
\newblock \bibinfo{journal}{\bibinfo{title}{{Mimetic gravity: a review of
  recent developments and applications to cosmology and astrophysics}}}.
\newblock {\emph{\JournalTitle{Adv. High Energy Phys.}}}
  \textbf{\bibinfo{volume}{2017}}, \bibinfo{pages}{3156915},
  \doiprefix\url{10.1155/2017/3156915} (\bibinfo{year}{2017}).
\newblock \eprint{1612.08661}.

\bibitem{Chamseddine:2014vna}
\bibinfo{author}{Chamseddine, A.~H.}, \bibinfo{author}{Mukhanov, V.} \&
  \bibinfo{author}{Vikman, A.}
\newblock \bibinfo{journal}{\bibinfo{title}{{Cosmology with Mimetic Matter}}}.
\newblock {\emph{\JournalTitle{JCAP}}} \textbf{\bibinfo{volume}{06}},
  \bibinfo{pages}{017}, \doiprefix\url{10.1088/1475-7516/2014/06/017}
  (\bibinfo{year}{2014}).
\newblock \eprint{1403.3961}.

\bibitem{Chaichian:2014qba}
\bibinfo{author}{Chaichian, M.}, \bibinfo{author}{Kluson, J.},
  \bibinfo{author}{Oksanen, M.} \& \bibinfo{author}{Tureanu, A.}
\newblock \bibinfo{journal}{\bibinfo{title}{{Mimetic dark matter, ghost
  instability and a mimetic tensor-vector-scalar gravity}}}.
\newblock {\emph{\JournalTitle{JHEP}}} \textbf{\bibinfo{volume}{12}},
  \bibinfo{pages}{102}, \doiprefix\url{10.1007/JHEP12(2014)102}
  (\bibinfo{year}{2014}).
\newblock \eprint{1404.4008}.

\bibitem{Nojiri:2014zqa}
\bibinfo{author}{Nojiri, S.} \& \bibinfo{author}{Odintsov, S.~D.}
\newblock \bibinfo{journal}{\bibinfo{title}{{Mimetic $F(R)$ gravity: inflation,
  dark energy and bounce}}}.
\newblock {\emph{\JournalTitle{Mod. Phys. Lett. A}}}
  \textbf{\bibinfo{volume}{29}}, \bibinfo{pages}{1450211},
  \doiprefix\url{10.1142/S0217732314502113} (\bibinfo{year}{2014}).
\newblock \eprint{1408.3561}.

\bibitem{Mirzagholi:2014ifa}
\bibinfo{author}{Mirzagholi, L.} \& \bibinfo{author}{Vikman, A.}
\newblock \bibinfo{journal}{\bibinfo{title}{{Imperfect Dark Matter}}}.
\newblock {\emph{\JournalTitle{JCAP}}} \textbf{\bibinfo{volume}{06}},
  \bibinfo{pages}{028}, \doiprefix\url{10.1088/1475-7516/2015/06/028}
  (\bibinfo{year}{2015}).
\newblock \eprint{1412.7136}.

\bibitem{Leon:2014yua}
\bibinfo{author}{Leon, G.} \& \bibinfo{author}{Saridakis, E.~N.}
\newblock \bibinfo{journal}{\bibinfo{title}{{Dynamical behavior in mimetic F(R)
  gravity}}}.
\newblock {\emph{\JournalTitle{JCAP}}} \textbf{\bibinfo{volume}{04}},
  \bibinfo{pages}{031}, \doiprefix\url{10.1088/1475-7516/2015/04/031}
  (\bibinfo{year}{2015}).
\newblock \eprint{1501.00488}.

\bibitem{Momeni:2015gka}
\bibinfo{author}{Momeni, D.}, \bibinfo{author}{Myrzakulov, R.} \&
  \bibinfo{author}{G\"udekli, E.}
\newblock \bibinfo{journal}{\bibinfo{title}{{Cosmological viable mimetic $f(R)$
  and $f(R,T)$ theories via Noether symmetry}}}.
\newblock {\emph{\JournalTitle{Int. J. Geom. Meth. Mod. Phys.}}}
  \textbf{\bibinfo{volume}{12}}, \bibinfo{pages}{1550101},
  \doiprefix\url{10.1142/S0219887815501017} (\bibinfo{year}{2015}).
\newblock \eprint{1502.00977}.

\bibitem{Myrzakulov:2015qaa}
\bibinfo{author}{Myrzakulov, R.}, \bibinfo{author}{Sebastiani, L.} \&
  \bibinfo{author}{Vagnozzi, S.}
\newblock \bibinfo{journal}{\bibinfo{title}{{Inflation in $f(R,\phi )$
  -theories and mimetic gravity scenario}}}.
\newblock {\emph{\JournalTitle{Eur. Phys. J. C}}}
  \textbf{\bibinfo{volume}{75}}, \bibinfo{pages}{444},
  \doiprefix\url{10.1140/epjc/s10052-015-3672-6} (\bibinfo{year}{2015}).
\newblock \eprint{1504.07984}.

\bibitem{Astashenok:2015haa}
\bibinfo{author}{Astashenok, A.~V.}, \bibinfo{author}{Odintsov, S.~D.} \&
  \bibinfo{author}{Oikonomou, V.~K.}
\newblock \bibinfo{journal}{\bibinfo{title}{{Modified Gauss\textendash{}Bonnet
  gravity with the Lagrange multiplier constraint as mimetic theory}}}.
\newblock {\emph{\JournalTitle{Class. Quant. Grav.}}}
  \textbf{\bibinfo{volume}{32}}, \bibinfo{pages}{185007},
  \doiprefix\url{10.1088/0264-9381/32/18/185007} (\bibinfo{year}{2015}).
\newblock \eprint{1504.04861}.

\bibitem{Rabochaya:2015haa}
\bibinfo{author}{Rabochaya, Y.} \& \bibinfo{author}{Zerbini, S.}
\newblock \bibinfo{journal}{\bibinfo{title}{{A note on a mimetic
  scalar\textendash{}tensor cosmological model}}}.
\newblock {\emph{\JournalTitle{Eur. Phys. J. C}}}
  \textbf{\bibinfo{volume}{76}}, \bibinfo{pages}{85},
  \doiprefix\url{10.1140/epjc/s10052-016-3926-y} (\bibinfo{year}{2016}).
\newblock \eprint{1509.03720}.

\bibitem{Odintsov:2016oyz}
\bibinfo{author}{Odintsov, S.~D.} \& \bibinfo{author}{Oikonomou, V.~K.}
\newblock \bibinfo{journal}{\bibinfo{title}{{Dark Energy Oscillations in
  Mimetic $F(R)$ Gravity}}}.
\newblock {\emph{\JournalTitle{Phys. Rev. D}}} \textbf{\bibinfo{volume}{94}},
  \bibinfo{pages}{044012}, \doiprefix\url{10.1103/PhysRevD.94.044012}
  (\bibinfo{year}{2016}).
\newblock \eprint{1608.00165}.

\bibitem{Oikonomou:2016pkp}
\bibinfo{author}{Oikonomou, V.~K.}
\newblock \bibinfo{journal}{\bibinfo{title}{{Aspects of Late-time Evolution in
  Mimetic $F(R)$ Gravity}}}.
\newblock {\emph{\JournalTitle{Mod. Phys. Lett. A}}}
  \textbf{\bibinfo{volume}{31}}, \bibinfo{pages}{1650191},
  \doiprefix\url{10.1142/S0217732316501911} (\bibinfo{year}{2016}).
\newblock \eprint{1609.03156}.

\bibitem{Firouzjahi:2017txv}
\bibinfo{author}{Firouzjahi, H.}, \bibinfo{author}{Gorji, M.~A.} \&
  \bibinfo{author}{Hosseini~Mansoori, S.~A.}
\newblock \bibinfo{journal}{\bibinfo{title}{{Instabilities in Mimetic Matter
  Perturbations}}}.
\newblock {\emph{\JournalTitle{JCAP}}} \textbf{\bibinfo{volume}{07}},
  \bibinfo{pages}{031}, \doiprefix\url{10.1088/1475-7516/2017/07/031}
  (\bibinfo{year}{2017}).
\newblock \eprint{1703.02923}.

\bibitem{Hirano:2017zox}
\bibinfo{author}{Hirano, S.}, \bibinfo{author}{Nishi, S.} \&
  \bibinfo{author}{Kobayashi, T.}
\newblock \bibinfo{journal}{\bibinfo{title}{{Healthy imperfect dark matter from
  effective theory of mimetic cosmological perturbations}}}.
\newblock {\emph{\JournalTitle{JCAP}}} \textbf{\bibinfo{volume}{07}},
  \bibinfo{pages}{009}, \doiprefix\url{10.1088/1475-7516/2017/07/009}
  (\bibinfo{year}{2017}).
\newblock \eprint{1704.06031}.

\bibitem{Zheng:2017qfs}
\bibinfo{author}{Zheng, Y.}, \bibinfo{author}{Shen, L.}, \bibinfo{author}{Mou,
  Y.} \& \bibinfo{author}{Li, M.}
\newblock \bibinfo{journal}{\bibinfo{title}{{On (in)stabilities of
  perturbations in mimetic models with higher derivatives}}}.
\newblock {\emph{\JournalTitle{JCAP}}} \textbf{\bibinfo{volume}{08}},
  \bibinfo{pages}{040}, \doiprefix\url{10.1088/1475-7516/2017/08/040}
  (\bibinfo{year}{2017}).
\newblock \eprint{1704.06834}.

\bibitem{Vagnozzi:2017ilo}
\bibinfo{author}{Vagnozzi, S.}
\newblock \bibinfo{journal}{\bibinfo{title}{{Recovering a MOND-like
  acceleration law in mimetic gravity}}}.
\newblock {\emph{\JournalTitle{Class. Quant. Grav.}}}
  \textbf{\bibinfo{volume}{34}}, \bibinfo{pages}{185006},
  \doiprefix\url{10.1088/1361-6382/aa838b} (\bibinfo{year}{2017}).
\newblock \eprint{1708.00603}.

\bibitem{Takahashi:2017pje}
\bibinfo{author}{Takahashi, K.} \& \bibinfo{author}{Kobayashi, T.}
\newblock \bibinfo{journal}{\bibinfo{title}{{Extended mimetic gravity:
  Hamiltonian analysis and gradient instabilities}}}.
\newblock {\emph{\JournalTitle{JCAP}}} \textbf{\bibinfo{volume}{11}},
  \bibinfo{pages}{038}, \doiprefix\url{10.1088/1475-7516/2017/11/038}
  (\bibinfo{year}{2017}).
\newblock \eprint{1708.02951}.

\bibitem{Gorji:2017cai}
\bibinfo{author}{Gorji, M.~A.}, \bibinfo{author}{Hosseini~Mansoori, S.~A.} \&
  \bibinfo{author}{Firouzjahi, H.}
\newblock \bibinfo{journal}{\bibinfo{title}{{Higher Derivative Mimetic
  Gravity}}}.
\newblock {\emph{\JournalTitle{JCAP}}} \textbf{\bibinfo{volume}{01}},
  \bibinfo{pages}{020}, \doiprefix\url{10.1088/1475-7516/2018/01/020}
  (\bibinfo{year}{2018}).
\newblock \eprint{1709.09988}.

\bibitem{Dutta:2017fjw}
\bibinfo{author}{Dutta, J.}, \bibinfo{author}{Khyllep, W.},
  \bibinfo{author}{Saridakis, E.~N.}, \bibinfo{author}{Tamanini, N.} \&
  \bibinfo{author}{Vagnozzi, S.}
\newblock \bibinfo{journal}{\bibinfo{title}{{Cosmological dynamics of mimetic
  gravity}}}.
\newblock {\emph{\JournalTitle{JCAP}}} \textbf{\bibinfo{volume}{02}},
  \bibinfo{pages}{041}, \doiprefix\url{10.1088/1475-7516/2018/02/041}
  (\bibinfo{year}{2018}).
\newblock \eprint{1711.07290}.

\bibitem{Odintsov:2018ggm}
\bibinfo{author}{Odintsov, S.~D.} \& \bibinfo{author}{Oikonomou, V.~K.}
\newblock \bibinfo{journal}{\bibinfo{title}{{The reconstruction of $f(\phi)R$
  and mimetic gravity from viable slow-roll inflation}}}.
\newblock {\emph{\JournalTitle{Nucl. Phys. B}}} \textbf{\bibinfo{volume}{929}},
  \bibinfo{pages}{79--112}, \doiprefix\url{10.1016/j.nuclphysb.2018.01.027}
  (\bibinfo{year}{2018}).
\newblock \eprint{1801.10529}.

\bibitem{Casalino:2018tcd}
\bibinfo{author}{Casalino, A.}, \bibinfo{author}{Rinaldi, M.},
  \bibinfo{author}{Sebastiani, L.} \& \bibinfo{author}{Vagnozzi, S.}
\newblock \bibinfo{journal}{\bibinfo{title}{{Mimicking dark matter and dark
  energy in a mimetic model compatible with GW170817}}}.
\newblock {\emph{\JournalTitle{Phys. Dark Univ.}}}
  \textbf{\bibinfo{volume}{22}}, \bibinfo{pages}{108},
  \doiprefix\url{10.1016/j.dark.2018.10.001} (\bibinfo{year}{2018}).
\newblock \eprint{1803.02620}.

\bibitem{Ganz:2018mqi}
\bibinfo{author}{Ganz, A.}, \bibinfo{author}{Karmakar, P.},
  \bibinfo{author}{Matarrese, S.} \& \bibinfo{author}{Sorokin, D.}
\newblock \bibinfo{journal}{\bibinfo{title}{{Hamiltonian analysis of mimetic
  scalar gravity revisited}}}.
\newblock {\emph{\JournalTitle{Phys. Rev. D}}} \textbf{\bibinfo{volume}{99}},
  \bibinfo{pages}{064009}, \doiprefix\url{10.1103/PhysRevD.99.064009}
  (\bibinfo{year}{2019}).
\newblock \eprint{1812.02667}.

\bibitem{Solomon:2019qgf}
\bibinfo{author}{Solomon, A.~R.}, \bibinfo{author}{Vardanyan, V.} \&
  \bibinfo{author}{Akrami, Y.}
\newblock \bibinfo{journal}{\bibinfo{title}{{Massive mimetic cosmology}}}.
\newblock {\emph{\JournalTitle{Phys. Lett. B}}} \textbf{\bibinfo{volume}{794}},
  \bibinfo{pages}{135--142}, \doiprefix\url{10.1016/j.physletb.2019.05.045}
  (\bibinfo{year}{2019}).
\newblock \eprint{1902.08533}.

\bibitem{Gorji:2019ttx}
\bibinfo{author}{Gorji, M.~A.}, \bibinfo{author}{Mukohyama, S.} \&
  \bibinfo{author}{Firouzjahi, H.}
\newblock \bibinfo{journal}{\bibinfo{title}{{Cosmology in Mimetic SU(2) Gauge
  Theory}}}.
\newblock {\emph{\JournalTitle{JCAP}}} \textbf{\bibinfo{volume}{05}},
  \bibinfo{pages}{019}, \doiprefix\url{10.1088/1475-7516/2019/05/019}
  (\bibinfo{year}{2019}).
\newblock \eprint{1903.04845}.

\bibitem{Khalifeh:2019zfi}
\bibinfo{author}{Khalifeh, A.~R.}, \bibinfo{author}{Bellomo, N.},
  \bibinfo{author}{Bernal, J.~L.} \& \bibinfo{author}{Jimenez, R.}
\newblock \bibinfo{journal}{\bibinfo{title}{{Can Dark Matter be Geometry? A
  Case Study with Mimetic Dark Matter}}}.
\newblock {\emph{\JournalTitle{Phys. Dark Univ.}}}
  \textbf{\bibinfo{volume}{30}}, \bibinfo{pages}{100646},
  \doiprefix\url{10.1016/j.dark.2020.100646} (\bibinfo{year}{2020}).
\newblock \eprint{1907.03660}.

\bibitem{Rashidi:2020jao}
\bibinfo{author}{Rashidi, N.} \& \bibinfo{author}{Nozari, K.}
\newblock \bibinfo{journal}{\bibinfo{title}{{Tachyon mimetic inflation as an
  instabilities-free model}}}.
\newblock {\emph{\JournalTitle{Phys. Rev. D}}} \textbf{\bibinfo{volume}{102}},
  \bibinfo{pages}{123548}, \doiprefix\url{10.1103/PhysRevD.102.123548}
  (\bibinfo{year}{2020}).
\newblock \eprint{2101.00825}.

\bibitem{Kaczmarek:2021psy}
\bibinfo{author}{Kaczmarek, A.~Z.} \& \bibinfo{author}{Szczesniak, D.}
\newblock \bibinfo{journal}{\bibinfo{title}{{Cosmology in the mimetic
  higher-curvature $f(R,R_{\mu \nu }R^{\mu \nu })$ gravity}}}.
\newblock {\emph{\JournalTitle{Sci. Rep.}}} \textbf{\bibinfo{volume}{11}},
  \bibinfo{pages}{18363}, \doiprefix\url{10.1038/s41598-021-97907-y}
  (\bibinfo{year}{2021}).
\newblock \eprint{2105.05050}.

\bibitem{Benisty:2021cin}
\bibinfo{author}{Benisty, D.}, \bibinfo{author}{Chaichian, M.~M.} \&
  \bibinfo{author}{Oksanen, M.}
\newblock \bibinfo{journal}{\bibinfo{title}{{Mimetic
  tensor\textendash{}vector\textendash{}scalar cosmology: Incorporating dark
  matter, dark energy and stiff matter}}}.
\newblock {\emph{\JournalTitle{Phys. Dark Univ.}}}
  \textbf{\bibinfo{volume}{42}}, \bibinfo{pages}{101280},
  \doiprefix\url{10.1016/j.dark.2023.101280} (\bibinfo{year}{2023}).
\newblock \eprint{2107.12161}.

\bibitem{Nashed:2022yfc}
\bibinfo{author}{Nashed, G. G.~L.} \& \bibinfo{author}{Saridakis, E.~N.}
\newblock \bibinfo{journal}{\bibinfo{title}{{New anisotropic star solutions in
  mimetic gravity}}}.
\newblock {\emph{\JournalTitle{Eur. Phys. J. Plus}}}
  \textbf{\bibinfo{volume}{138}}, \bibinfo{pages}{318},
  \doiprefix\url{10.1140/epjp/s13360-023-03767-y} (\bibinfo{year}{2023}).
\newblock \eprint{2206.12256}.

\bibitem{Nashed:2023jdf}
\bibinfo{author}{Nashed, G. G.~L.}
\newblock \bibinfo{journal}{\bibinfo{title}{{The key role of Lagrangian
  multiplier in mimetic gravitational theory in the frame of isotropic compact
  star}}}.
\newblock {\emph{\JournalTitle{Nucl. Phys. B}}} \textbf{\bibinfo{volume}{993}},
  \bibinfo{pages}{116264}, \doiprefix\url{10.1016/j.nuclphysb.2023.116264}
  (\bibinfo{year}{2023}).
\newblock \eprint{2307.03199}.

\bibitem{Kaczmarek:2023qmq}
\bibinfo{author}{Kaczmarek, A.~Z.} \& \bibinfo{author}{Szczesniak, D.}
\newblock \bibinfo{journal}{\bibinfo{title}{{Cosmological aspects of the
  unimodular-mimetic f(G) gravity}}}.
\newblock {\emph{\JournalTitle{Nucl. Phys. B}}}
  \textbf{\bibinfo{volume}{1002}}, \bibinfo{pages}{116534},
  \doiprefix\url{10.1016/j.nuclphysb.2024.116534} (\bibinfo{year}{2024}).
\newblock \eprint{2311.05960}.

\bibitem{Casalino:2018wnc}
\bibinfo{author}{Casalino, A.}, \bibinfo{author}{Rinaldi, M.},
  \bibinfo{author}{Sebastiani, L.} \& \bibinfo{author}{Vagnozzi, S.}
\newblock \bibinfo{journal}{\bibinfo{title}{{Alive and well: mimetic gravity
  and a higher-order extension in light of GW170817}}}.
\newblock {\emph{\JournalTitle{Class. Quant. Grav.}}}
  \textbf{\bibinfo{volume}{36}}, \bibinfo{pages}{017001},
  \doiprefix\url{10.1088/1361-6382/aaf1fd} (\bibinfo{year}{2019}).
\newblock \eprint{1811.06830}.

\bibitem{Gorji:2020ten}
\bibinfo{author}{Gorji, M.~A.}, \bibinfo{author}{Allahyari, A.},
  \bibinfo{author}{Khodadi, M.} \& \bibinfo{author}{Firouzjahi, H.}
\newblock \bibinfo{journal}{\bibinfo{title}{{Mimetic black holes}}}.
\newblock {\emph{\JournalTitle{Phys. Rev. D}}} \textbf{\bibinfo{volume}{101}},
  \bibinfo{pages}{124060}, \doiprefix\url{10.1103/PhysRevD.101.124060}
  (\bibinfo{year}{2020}).
\newblock \eprint{1912.04636}.

\bibitem{EventHorizonTelescope:2019dse}
\bibinfo{author}{Akiyama, K.} \emph{et~al.}
\newblock \bibinfo{journal}{\bibinfo{title}{{First M87 Event Horizon Telescope
  Results. I. The Shadow of the Supermassive Black Hole}}}.
\newblock {\emph{\JournalTitle{Astrophys. J. Lett.}}}
  \textbf{\bibinfo{volume}{875}}, \bibinfo{pages}{L1},
  \doiprefix\url{10.3847/2041-8213/ab0ec7} (\bibinfo{year}{2019}).
\newblock \eprint{1906.11238}.

\bibitem{EventHorizonTelescope:2022wkp}
\bibinfo{author}{Akiyama, K.} \emph{et~al.}
\newblock \bibinfo{journal}{\bibinfo{title}{{First Sagittarius A* Event Horizon
  Telescope Results. I. The Shadow of the Supermassive Black Hole in the Center
  of the Milky Way}}}.
\newblock {\emph{\JournalTitle{Astrophys. J. Lett.}}}
  \textbf{\bibinfo{volume}{930}}, \bibinfo{pages}{L12},
  \doiprefix\url{10.3847/2041-8213/ac6674} (\bibinfo{year}{2022}).
\newblock \eprint{2311.08680}.

\bibitem{Zumalacarregui:2013pma}
\bibinfo{author}{Zumalac\'arregui, M.} \& \bibinfo{author}{Garc\'\i{}a-Bellido,
  J.}
\newblock \bibinfo{journal}{\bibinfo{title}{{Transforming gravity: from
  derivative couplings to matter to second-order scalar-tensor theories beyond
  the Horndeski Lagrangian}}}.
\newblock {\emph{\JournalTitle{Phys. Rev. D}}} \textbf{\bibinfo{volume}{89}},
  \bibinfo{pages}{064046}, \doiprefix\url{10.1103/PhysRevD.89.064046}
  (\bibinfo{year}{2014}).
\newblock \eprint{1308.4685}.

\bibitem{Deruelle:2014zza}
\bibinfo{author}{Deruelle, N.} \& \bibinfo{author}{Rua, J.}
\newblock \bibinfo{journal}{\bibinfo{title}{{Disformal Transformations, Veiled
  General Relativity and Mimetic Gravity}}}.
\newblock {\emph{\JournalTitle{JCAP}}} \textbf{\bibinfo{volume}{09}},
  \bibinfo{pages}{002}, \doiprefix\url{10.1088/1475-7516/2014/09/002}
  (\bibinfo{year}{2014}).
\newblock \eprint{1407.0825}.

\bibitem{Domenech:2015tca}
\bibinfo{author}{Dom\`enech, G.} \emph{et~al.}
\newblock \bibinfo{journal}{\bibinfo{title}{{Derivative-dependent metric
  transformation and physical degrees of freedom}}}.
\newblock {\emph{\JournalTitle{Phys. Rev. D}}} \textbf{\bibinfo{volume}{92}},
  \bibinfo{pages}{084027}, \doiprefix\url{10.1103/PhysRevD.92.084027}
  (\bibinfo{year}{2015}).
\newblock \eprint{1507.05390}.

\bibitem{Arroja:2015wpa}
\bibinfo{author}{Arroja, F.}, \bibinfo{author}{Bartolo, N.},
  \bibinfo{author}{Karmakar, P.} \& \bibinfo{author}{Matarrese, S.}
\newblock \bibinfo{journal}{\bibinfo{title}{{The two faces of mimetic Horndeski
  gravity: disformal transformations and Lagrange multiplier}}}.
\newblock {\emph{\JournalTitle{JCAP}}} \textbf{\bibinfo{volume}{09}},
  \bibinfo{pages}{051}, \doiprefix\url{10.1088/1475-7516/2015/09/051}
  (\bibinfo{year}{2015}).
\newblock \eprint{1506.08575}.

\bibitem{BenAchour:2016cay}
\bibinfo{author}{Ben~Achour, J.}, \bibinfo{author}{Langlois, D.} \&
  \bibinfo{author}{Noui, K.}
\newblock \bibinfo{journal}{\bibinfo{title}{{Degenerate higher order
  scalar-tensor theories beyond Horndeski and disformal transformations}}}.
\newblock {\emph{\JournalTitle{Phys. Rev. D}}} \textbf{\bibinfo{volume}{93}},
  \bibinfo{pages}{124005}, \doiprefix\url{10.1103/PhysRevD.93.124005}
  (\bibinfo{year}{2016}).
\newblock \eprint{1602.08398}.

\bibitem{Jirousek:2022rym}
\bibinfo{author}{Jirou\v{s}ek, P.}, \bibinfo{author}{Shimada, K.},
  \bibinfo{author}{Vikman, A.} \& \bibinfo{author}{Yamaguchi, M.}
\newblock \bibinfo{journal}{\bibinfo{title}{{Disforming to conformal
  symmetry}}}.
\newblock {\emph{\JournalTitle{JCAP}}} \textbf{\bibinfo{volume}{11}},
  \bibinfo{pages}{019}, \doiprefix\url{10.1088/1475-7516/2022/11/019}
  (\bibinfo{year}{2022}).
\newblock \eprint{2207.12611}.

\bibitem{Domenech:2023ryc}
\bibinfo{author}{Dom\`enech, G.} \& \bibinfo{author}{Ganz, A.}
\newblock \bibinfo{journal}{\bibinfo{title}{{Disformal symmetry in the
  Universe: mimetic gravity and beyond}}}.
\newblock {\emph{\JournalTitle{JCAP}}} \textbf{\bibinfo{volume}{08}},
  \bibinfo{pages}{046}, \doiprefix\url{10.1088/1475-7516/2023/08/046}
  (\bibinfo{year}{2023}).
\newblock \eprint{2304.11035}.

\bibitem{Golovnev:2013jxa}
\bibinfo{author}{Golovnev, A.}
\newblock \bibinfo{journal}{\bibinfo{title}{{On the recently proposed Mimetic
  Dark Matter}}}.
\newblock {\emph{\JournalTitle{Phys. Lett. B}}} \textbf{\bibinfo{volume}{728}},
  \bibinfo{pages}{39--40}, \doiprefix\url{10.1016/j.physletb.2013.11.026}
  (\bibinfo{year}{2014}).
\newblock \eprint{1310.2790}.

\bibitem{Lim:2010yk}
\bibinfo{author}{Lim, E.~A.}, \bibinfo{author}{Sawicki, I.} \&
  \bibinfo{author}{Vikman, A.}
\newblock \bibinfo{journal}{\bibinfo{title}{{Dust of Dark Energy}}}.
\newblock {\emph{\JournalTitle{JCAP}}} \textbf{\bibinfo{volume}{05}},
  \bibinfo{pages}{012}, \doiprefix\url{10.1088/1475-7516/2010/05/012}
  (\bibinfo{year}{2010}).
\newblock \eprint{1003.5751}.

\bibitem{Gao:2010gj}
\bibinfo{author}{Gao, C.}, \bibinfo{author}{Gong, Y.}, \bibinfo{author}{Wang,
  X.} \& \bibinfo{author}{Chen, X.}
\newblock \bibinfo{journal}{\bibinfo{title}{{Cosmological models with Lagrange
  Multiplier Field}}}.
\newblock {\emph{\JournalTitle{Phys. Lett. B}}} \textbf{\bibinfo{volume}{702}},
  \bibinfo{pages}{107--113}, \doiprefix\url{10.1016/j.physletb.2011.06.085}
  (\bibinfo{year}{2011}).
\newblock \eprint{1003.6056}.

\bibitem{Capozziello:2010uv}
\bibinfo{author}{Capozziello, S.}, \bibinfo{author}{Matsumoto, J.},
  \bibinfo{author}{Nojiri, S.} \& \bibinfo{author}{Odintsov, S.~D.}
\newblock \bibinfo{journal}{\bibinfo{title}{{Dark energy from modified gravity
  with Lagrange multipliers}}}.
\newblock {\emph{\JournalTitle{Phys. Lett. B}}} \textbf{\bibinfo{volume}{693}},
  \bibinfo{pages}{198--208}, \doiprefix\url{10.1016/j.physletb.2010.08.030}
  (\bibinfo{year}{2010}).
\newblock \eprint{1004.3691}.

\bibitem{Oikonomou:2016fxb}
\bibinfo{author}{Oikonomou, V.~K.}
\newblock \bibinfo{journal}{\bibinfo{title}{{A note on
  Schwarzschild\textendash{}de Sitter black holes in mimetic F(R) gravity}}}.
\newblock {\emph{\JournalTitle{Int. J. Mod. Phys. D}}}
  \textbf{\bibinfo{volume}{25}}, \bibinfo{pages}{1650078},
  \doiprefix\url{10.1142/S0218271816500784} (\bibinfo{year}{2016}).
\newblock \eprint{1605.00583}.

\bibitem{Myrzakulov:2015sea}
\bibinfo{author}{Myrzakulov, R.} \& \bibinfo{author}{Sebastiani, L.}
\newblock \bibinfo{journal}{\bibinfo{title}{{Spherically symmetric static
  vacuum solutions in Mimetic gravity}}}.
\newblock {\emph{\JournalTitle{Gen. Rel. Grav.}}}
  \textbf{\bibinfo{volume}{47}}, \bibinfo{pages}{89},
  \doiprefix\url{10.1007/s10714-015-1930-4} (\bibinfo{year}{2015}).
\newblock \eprint{1503.04293}.

\bibitem{Myrzakulov:2015kda}
\bibinfo{author}{Myrzakulov, R.}, \bibinfo{author}{Sebastiani, L.},
  \bibinfo{author}{Vagnozzi, S.} \& \bibinfo{author}{Zerbini, S.}
\newblock \bibinfo{journal}{\bibinfo{title}{{Static spherically symmetric
  solutions in mimetic gravity: rotation curves and wormholes}}}.
\newblock {\emph{\JournalTitle{Class. Quant. Grav.}}}
  \textbf{\bibinfo{volume}{33}}, \bibinfo{pages}{125005},
  \doiprefix\url{10.1088/0264-9381/33/12/125005} (\bibinfo{year}{2016}).
\newblock \eprint{1510.02284}.

\bibitem{Sheykhi:2019gvk}
\bibinfo{author}{Sheykhi, A.} \& \bibinfo{author}{Grunau, S.}
\newblock \bibinfo{journal}{\bibinfo{title}{{Topological black holes in mimetic
  gravity}}}.
\newblock {\emph{\JournalTitle{Int. J. Mod. Phys. A}}}
  \textbf{\bibinfo{volume}{36}}, \bibinfo{pages}{2150186},
  \doiprefix\url{10.1142/S0217751X21501864} (\bibinfo{year}{2021}).
\newblock \eprint{1911.13072}.

\bibitem{Izumi:2009ry}
\bibinfo{author}{Izumi, K.} \& \bibinfo{author}{Mukohyama, S.}
\newblock \bibinfo{journal}{\bibinfo{title}{{Stellar center is dynamical in
  Horava-Lifshitz gravity}}}.
\newblock {\emph{\JournalTitle{Phys. Rev. D}}} \textbf{\bibinfo{volume}{81}},
  \bibinfo{pages}{044008}, \doiprefix\url{10.1103/PhysRevD.81.044008}
  (\bibinfo{year}{2010}).
\newblock \eprint{0911.1814}.

\bibitem{Cognola:2016gjy}
\bibinfo{author}{Cognola, G.}, \bibinfo{author}{Myrzakulov, R.},
  \bibinfo{author}{Sebastiani, L.}, \bibinfo{author}{Vagnozzi, S.} \&
  \bibinfo{author}{Zerbini, S.}
\newblock \bibinfo{journal}{\bibinfo{title}{{Covariant Ho\v{r}ava-like and
  mimetic Horndeski gravity: cosmological solutions and perturbations}}}.
\newblock {\emph{\JournalTitle{Class. Quant. Grav.}}}
  \textbf{\bibinfo{volume}{33}}, \bibinfo{pages}{225014},
  \doiprefix\url{10.1088/0264-9381/33/22/225014} (\bibinfo{year}{2016}).
\newblock \eprint{1601.00102}.

\bibitem{Ramazanov:2016xhp}
\bibinfo{author}{Ramazanov, S.}, \bibinfo{author}{Arroja, F.},
  \bibinfo{author}{Celoria, M.}, \bibinfo{author}{Matarrese, S.} \&
  \bibinfo{author}{Pilo, L.}
\newblock \bibinfo{journal}{\bibinfo{title}{{Living with ghosts in
  Ho\v{r}ava-Lifshitz gravity}}}.
\newblock {\emph{\JournalTitle{JHEP}}} \textbf{\bibinfo{volume}{06}},
  \bibinfo{pages}{020}, \doiprefix\url{10.1007/JHEP06(2016)020}
  (\bibinfo{year}{2016}).
\newblock \eprint{1601.05405}.

\bibitem{Chamseddine:2019gjh}
\bibinfo{author}{Chamseddine, A.~H.}, \bibinfo{author}{Mukhanov, V.} \&
  \bibinfo{author}{Russ, T.~B.}
\newblock \bibinfo{journal}{\bibinfo{title}{{Mimetic Ho\v{r}ava gravity}}}.
\newblock {\emph{\JournalTitle{Phys. Lett. B}}} \textbf{\bibinfo{volume}{798}},
  \bibinfo{pages}{134939}, \doiprefix\url{10.1016/j.physletb.2019.134939}
  (\bibinfo{year}{2019}).
\newblock \eprint{1908.01717}.

\bibitem{Nashed:2018qag}
\bibinfo{author}{Nashed, G. G.~L.}, \bibinfo{author}{El~Hanafy, W.} \&
  \bibinfo{author}{Bamba, K.}
\newblock \bibinfo{journal}{\bibinfo{title}{{Charged rotating black holes
  coupled with nonlinear electrodynamics Maxwell field in the mimetic
  gravity}}}.
\newblock {\emph{\JournalTitle{JCAP}}} \textbf{\bibinfo{volume}{01}},
  \bibinfo{pages}{058}, \doiprefix\url{10.1088/1475-7516/2019/01/058}
  (\bibinfo{year}{2019}).
\newblock \eprint{1809.02289}.

\bibitem{Nashed:2021ctg}
\bibinfo{author}{Nashed, G. G.~L.} \& \bibinfo{author}{Nojiri, S.}
\newblock \bibinfo{journal}{\bibinfo{title}{{Mimetic Euler-Heisenberg theory,
  charged solutions, and multihorizon black holes}}}.
\newblock {\emph{\JournalTitle{Phys. Rev. D}}} \textbf{\bibinfo{volume}{104}},
  \bibinfo{pages}{044043}, \doiprefix\url{10.1103/PhysRevD.104.044043}
  (\bibinfo{year}{2021}).
\newblock \eprint{2107.13550}.

\bibitem{Nashed:2021hgn}
\bibinfo{author}{Nashed, G. G.~L.} \& \bibinfo{author}{Nojiri, S.}
\newblock \bibinfo{journal}{\bibinfo{title}{{Black holes with Lagrange
  multiplier and potential in mimetic-like gravitational theory: multi-horizon
  black holes}}}.
\newblock {\emph{\JournalTitle{JCAP}}} \textbf{\bibinfo{volume}{05}},
  \bibinfo{pages}{011}, \doiprefix\url{10.1088/1475-7516/2022/05/011}
  (\bibinfo{year}{2022}).
\newblock \eprint{2110.08560}.

\bibitem{Cunha:2018acu}
\bibinfo{author}{Cunha, P. V.~P.} \& \bibinfo{author}{Herdeiro, C. A.~R.}
\newblock \bibinfo{journal}{\bibinfo{title}{{Shadows and strong gravitational
  lensing: a brief review}}}.
\newblock {\emph{\JournalTitle{Gen. Rel. Grav.}}}
  \textbf{\bibinfo{volume}{50}}, \bibinfo{pages}{42},
  \doiprefix\url{10.1007/s10714-018-2361-9} (\bibinfo{year}{2018}).
\newblock \eprint{1801.00860}.

\bibitem{Perlick:2021aok}
\bibinfo{author}{Perlick, V.} \& \bibinfo{author}{Tsupko, O.~Y.}
\newblock \bibinfo{journal}{\bibinfo{title}{{Calculating black hole shadows:
  Review of analytical studies}}}.
\newblock {\emph{\JournalTitle{Phys. Rept.}}} \textbf{\bibinfo{volume}{947}},
  \bibinfo{pages}{1--39}, \doiprefix\url{10.1016/j.physrep.2021.10.004}
  (\bibinfo{year}{2022}).
\newblock \eprint{2105.07101}.

\bibitem{Chen:2022scf}
\bibinfo{author}{Chen, S.}, \bibinfo{author}{Jing, J.}, \bibinfo{author}{Qian,
  W.-L.} \& \bibinfo{author}{Wang, B.}
\newblock \bibinfo{journal}{\bibinfo{title}{{Black hole images: A review}}}.
\newblock {\emph{\JournalTitle{Sci. China Phys. Mech. Astron.}}}
  \textbf{\bibinfo{volume}{66}}, \bibinfo{pages}{260401},
  \doiprefix\url{10.1007/s11433-022-2059-5} (\bibinfo{year}{2023}).
\newblock \eprint{2301.00113}.

\bibitem{Falcke:1999pj}
\bibinfo{author}{Falcke, H.}, \bibinfo{author}{Melia, F.} \&
  \bibinfo{author}{Agol, E.}
\newblock \bibinfo{journal}{\bibinfo{title}{{Viewing the shadow of the black
  hole at the galactic center}}}.
\newblock {\emph{\JournalTitle{Astrophys. J. Lett.}}}
  \textbf{\bibinfo{volume}{528}}, \bibinfo{pages}{L13},
  \doiprefix\url{10.1086/312423} (\bibinfo{year}{2000}).
\newblock \eprint{astro-ph/9912263}.

\bibitem{Held:2019xde}
\bibinfo{author}{Held, A.}, \bibinfo{author}{Gold, R.} \&
  \bibinfo{author}{Eichhorn, A.}
\newblock \bibinfo{journal}{\bibinfo{title}{{Asymptotic safety casts its
  shadow}}}.
\newblock {\emph{\JournalTitle{JCAP}}} \textbf{\bibinfo{volume}{06}},
  \bibinfo{pages}{029}, \doiprefix\url{10.1088/1475-7516/2019/06/029}
  (\bibinfo{year}{2019}).
\newblock \eprint{1904.07133}.

\bibitem{Vagnozzi:2019apd}
\bibinfo{author}{Vagnozzi, S.} \& \bibinfo{author}{Visinelli, L.}
\newblock \bibinfo{journal}{\bibinfo{title}{{Hunting for extra dimensions in
  the shadow of M87*}}}.
\newblock {\emph{\JournalTitle{Phys. Rev. D}}} \textbf{\bibinfo{volume}{100}},
  \bibinfo{pages}{024020}, \doiprefix\url{10.1103/PhysRevD.100.024020}
  (\bibinfo{year}{2019}).
\newblock \eprint{1905.12421}.

\bibitem{Zhu:2019ura}
\bibinfo{author}{Zhu, T.}, \bibinfo{author}{Wu, Q.}, \bibinfo{author}{Jamil,
  M.} \& \bibinfo{author}{Jusufi, K.}
\newblock \bibinfo{journal}{\bibinfo{title}{{Shadows and deflection angle of
  charged and slowly rotating black holes in Einstein-\AE{}ther theory}}}.
\newblock {\emph{\JournalTitle{Phys. Rev. D}}} \textbf{\bibinfo{volume}{100}},
  \bibinfo{pages}{044055}, \doiprefix\url{10.1103/PhysRevD.100.044055}
  (\bibinfo{year}{2019}).
\newblock \eprint{1906.05673}.

\bibitem{Cunha:2019ikd}
\bibinfo{author}{Cunha, P. V.~P.}, \bibinfo{author}{Herdeiro, C. A.~R.} \&
  \bibinfo{author}{Radu, E.}
\newblock \bibinfo{journal}{\bibinfo{title}{{EHT constraint on the ultralight
  scalar hair of the M87 supermassive black hole}}}.
\newblock {\emph{\JournalTitle{Universe}}} \textbf{\bibinfo{volume}{5}},
  \bibinfo{pages}{220}, \doiprefix\url{10.3390/universe5120220}
  (\bibinfo{year}{2019}).
\newblock \eprint{1909.08039}.

\bibitem{Banerjee:2019nnj}
\bibinfo{author}{Banerjee, I.}, \bibinfo{author}{Chakraborty, S.} \&
  \bibinfo{author}{SenGupta, S.}
\newblock \bibinfo{journal}{\bibinfo{title}{{Silhouette of M87*: A New Window
  to Peek into the World of Hidden Dimensions}}}.
\newblock {\emph{\JournalTitle{Phys. Rev. D}}} \textbf{\bibinfo{volume}{101}},
  \bibinfo{pages}{041301}, \doiprefix\url{10.1103/PhysRevD.101.041301}
  (\bibinfo{year}{2020}).
\newblock \eprint{1909.09385}.

\bibitem{Banerjee:2019xds}
\bibinfo{author}{Banerjee, I.}, \bibinfo{author}{Sau, S.} \&
  \bibinfo{author}{SenGupta, S.}
\newblock \bibinfo{journal}{\bibinfo{title}{{Implications of axionic hair on
  the shadow of M87*}}}.
\newblock {\emph{\JournalTitle{Phys. Rev. D}}} \textbf{\bibinfo{volume}{101}},
  \bibinfo{pages}{104057}, \doiprefix\url{10.1103/PhysRevD.101.104057}
  (\bibinfo{year}{2020}).
\newblock \eprint{1911.05385}.

\bibitem{Zhdanov:2019ozq}
\bibinfo{author}{Zhdanov, V.~I.} \& \bibinfo{author}{Stashko, O.~S.}
\newblock \bibinfo{journal}{\bibinfo{title}{{Static spherically symmetric
  configurations with N nonlinear scalar fields: Global and asymptotic
  properties}}}.
\newblock {\emph{\JournalTitle{Phys. Rev. D}}} \textbf{\bibinfo{volume}{101}},
  \bibinfo{pages}{064064}, \doiprefix\url{10.1103/PhysRevD.101.064064}
  (\bibinfo{year}{2020}).
\newblock \eprint{1912.00470}.

\bibitem{Allahyari:2019jqz}
\bibinfo{author}{Allahyari, A.}, \bibinfo{author}{Khodadi, M.},
  \bibinfo{author}{Vagnozzi, S.} \& \bibinfo{author}{Mota, D.~F.}
\newblock \bibinfo{journal}{\bibinfo{title}{{Magnetically charged black holes
  from non-linear electrodynamics and the Event Horizon Telescope}}}.
\newblock {\emph{\JournalTitle{JCAP}}} \textbf{\bibinfo{volume}{02}},
  \bibinfo{pages}{003}, \doiprefix\url{10.1088/1475-7516/2020/02/003}
  (\bibinfo{year}{2020}).
\newblock \eprint{1912.08231}.

\bibitem{Khodadi:2020jij}
\bibinfo{author}{Khodadi, M.}, \bibinfo{author}{Allahyari, A.},
  \bibinfo{author}{Vagnozzi, S.} \& \bibinfo{author}{Mota, D.~F.}
\newblock \bibinfo{journal}{\bibinfo{title}{{Black holes with scalar hair in
  light of the Event Horizon Telescope}}}.
\newblock {\emph{\JournalTitle{JCAP}}} \textbf{\bibinfo{volume}{09}},
  \bibinfo{pages}{026}, \doiprefix\url{10.1088/1475-7516/2020/09/026}
  (\bibinfo{year}{2020}).
\newblock \eprint{2005.05992}.

\bibitem{Kumar:2020yem}
\bibinfo{author}{Kumar, R.}, \bibinfo{author}{Kumar, A.} \&
  \bibinfo{author}{Ghosh, S.~G.}
\newblock \bibinfo{journal}{\bibinfo{title}{{Testing Rotating Regular Metrics
  as Candidates for Astrophysical Black Holes}}}.
\newblock {\emph{\JournalTitle{Astrophys. J.}}} \textbf{\bibinfo{volume}{896}},
  \bibinfo{pages}{89}, \doiprefix\url{10.3847/1538-4357/ab8c4a}
  (\bibinfo{year}{2020}).
\newblock \eprint{2006.09869}.

\bibitem{Khodadi:2020gns}
\bibinfo{author}{Khodadi, M.} \& \bibinfo{author}{Saridakis, E.~N.}
\newblock \bibinfo{journal}{\bibinfo{title}{{Einstein-\AE{}ther gravity in the
  light of event horizon telescope observations of M87*}}}.
\newblock {\emph{\JournalTitle{Phys. Dark Univ.}}}
  \textbf{\bibinfo{volume}{32}}, \bibinfo{pages}{100835},
  \doiprefix\url{10.1016/j.dark.2021.100835} (\bibinfo{year}{2021}).
\newblock \eprint{2012.05186}.

\bibitem{Pantig:2021zqe}
\bibinfo{author}{Pantig, R.~C.}, \bibinfo{author}{Yu, P.~K.},
  \bibinfo{author}{Rodulfo, E.~T.} \& \bibinfo{author}{\"Ovg\"un, A.}
\newblock \bibinfo{journal}{\bibinfo{title}{{Shadow and weak deflection angle
  of extended uncertainty principle black hole surrounded with dark matter}}}.
\newblock {\emph{\JournalTitle{Annals Phys.}}} \textbf{\bibinfo{volume}{436}},
  \bibinfo{pages}{168722}, \doiprefix\url{10.1016/j.aop.2021.168722}
  (\bibinfo{year}{2022}).
\newblock \eprint{2104.04304}.

\bibitem{EventHorizonTelescope:2021dqv}
\bibinfo{author}{Kocherlakota, P.} \emph{et~al.}
\newblock \bibinfo{journal}{\bibinfo{title}{{Constraints on black-hole charges
  with the 2017 EHT observations of M87*}}}.
\newblock {\emph{\JournalTitle{Phys. Rev. D}}} \textbf{\bibinfo{volume}{103}},
  \bibinfo{pages}{104047}, \doiprefix\url{10.1103/PhysRevD.103.104047}
  (\bibinfo{year}{2021}).
\newblock \eprint{2105.09343}.

\bibitem{Khodadi:2021gbc}
\bibinfo{author}{Khodadi, M.}, \bibinfo{author}{Lambiase, G.} \&
  \bibinfo{author}{Mota, D.~F.}
\newblock \bibinfo{journal}{\bibinfo{title}{{No-hair theorem in the wake of
  Event Horizon Telescope}}}.
\newblock {\emph{\JournalTitle{JCAP}}} \textbf{\bibinfo{volume}{09}},
  \bibinfo{pages}{028}, \doiprefix\url{10.1088/1475-7516/2021/09/028}
  (\bibinfo{year}{2021}).
\newblock \eprint{2107.00834}.

\bibitem{Stashko:2021lad}
\bibinfo{author}{Stashko, O.~S.}, \bibinfo{author}{Zhdanov, V.~I.} \&
  \bibinfo{author}{Alexandrov, A.~N.}
\newblock \bibinfo{journal}{\bibinfo{title}{{Thin accretion discs around
  spherically symmetric configurations with nonlinear scalar fields}}}.
\newblock {\emph{\JournalTitle{Phys. Rev. D}}} \textbf{\bibinfo{volume}{104}},
  \bibinfo{pages}{104055}, \doiprefix\url{10.1103/PhysRevD.104.104055}
  (\bibinfo{year}{2021}).
\newblock \eprint{2107.05111}.

\bibitem{Uniyal:2022vdu}
\bibinfo{author}{Uniyal, A.}, \bibinfo{author}{Pantig, R.~C.} \&
  \bibinfo{author}{\"Ovg\"un, A.}
\newblock \bibinfo{journal}{\bibinfo{title}{{Probing a non-linear
  electrodynamics black hole with thin accretion disk, shadow, and deflection
  angle with M87* and Sgr A* from EHT}}}.
\newblock {\emph{\JournalTitle{Phys. Dark Univ.}}}
  \textbf{\bibinfo{volume}{40}}, \bibinfo{pages}{101178},
  \doiprefix\url{10.1016/j.dark.2023.101178} (\bibinfo{year}{2023}).
\newblock \eprint{2205.11072}.

\bibitem{Pantig:2022ely}
\bibinfo{author}{Pantig, R.~C.} \& \bibinfo{author}{\"Ovg\"un, A.}
\newblock \bibinfo{journal}{\bibinfo{title}{{Testing dynamical torsion effects
  on the charged black hole\textquoteright{}s shadow, deflection angle and
  greybody with M87* and Sgr. A* from EHT}}}.
\newblock {\emph{\JournalTitle{Annals Phys.}}} \textbf{\bibinfo{volume}{448}},
  \bibinfo{pages}{169197}, \doiprefix\url{10.1016/j.aop.2022.169197}
  (\bibinfo{year}{2023}).
\newblock \eprint{2206.02161}.

\bibitem{Ghosh:2022kit}
\bibinfo{author}{Ghosh, S.~G.} \& \bibinfo{author}{Afrin, M.}
\newblock \bibinfo{journal}{\bibinfo{title}{{An Upper Limit on the Charge of
  the Black Hole Sgr A* from EHT Observations}}}.
\newblock {\emph{\JournalTitle{Astrophys. J.}}} \textbf{\bibinfo{volume}{944}},
  \bibinfo{pages}{174}, \doiprefix\url{10.3847/1538-4357/acb695}
  (\bibinfo{year}{2023}).
\newblock \eprint{2206.02488}.

\bibitem{Khodadi:2022pqh}
\bibinfo{author}{Khodadi, M.} \& \bibinfo{author}{Lambiase, G.}
\newblock \bibinfo{journal}{\bibinfo{title}{{Probing Lorentz symmetry violation
  using the first image of Sagittarius A*: Constraints on standard-model
  extension coefficients}}}.
\newblock {\emph{\JournalTitle{Phys. Rev. D}}} \textbf{\bibinfo{volume}{106}},
  \bibinfo{pages}{104050}, \doiprefix\url{10.1103/PhysRevD.106.104050}
  (\bibinfo{year}{2022}).
\newblock \eprint{2206.08601}.

\bibitem{KumarWalia:2022aop}
\bibinfo{author}{Kumar~Walia, R.}, \bibinfo{author}{Ghosh, S.~G.} \&
  \bibinfo{author}{Maharaj, S.~D.}
\newblock \bibinfo{journal}{\bibinfo{title}{{Testing Rotating Regular Metrics
  with EHT Results of Sgr A*}}}.
\newblock {\emph{\JournalTitle{Astrophys. J.}}} \textbf{\bibinfo{volume}{939}},
  \bibinfo{pages}{77}, \doiprefix\url{10.3847/1538-4357/ac9623}
  (\bibinfo{year}{2022}).
\newblock \eprint{2207.00078}.

\bibitem{Shaikh:2022ivr}
\bibinfo{author}{Shaikh, R.}
\newblock \bibinfo{journal}{\bibinfo{title}{{Testing black hole mimickers with
  the Event Horizon Telescope image of Sagittarius A*}}}.
\newblock {\emph{\JournalTitle{Mon. Not. Roy. Astron. Soc.}}}
  \textbf{\bibinfo{volume}{523}}, \bibinfo{pages}{375--384},
  \doiprefix\url{10.1093/mnras/stad1383} (\bibinfo{year}{2023}).
\newblock \eprint{2208.01995}.

\bibitem{Afrin:2022ztr}
\bibinfo{author}{Afrin, M.}, \bibinfo{author}{Vagnozzi, S.} \&
  \bibinfo{author}{Ghosh, S.~G.}
\newblock \bibinfo{journal}{\bibinfo{title}{{Tests of Loop Quantum Gravity from
  the Event Horizon Telescope Results of Sgr A*}}}.
\newblock {\emph{\JournalTitle{Astrophys. J.}}} \textbf{\bibinfo{volume}{944}},
  \bibinfo{pages}{149}, \doiprefix\url{10.3847/1538-4357/acb334}
  (\bibinfo{year}{2023}).
\newblock \eprint{2209.12584}.

\bibitem{Pantig:2023yer}
\bibinfo{author}{Pantig, R.~C.}
\newblock \bibinfo{journal}{\bibinfo{title}{{Constraining a one-dimensional
  wave-type gravitational wave parameter through the shadow of M87* via Event
  Horizon Telescope}}}.
\newblock {\emph{\JournalTitle{Chin. J. Phys.}}} \textbf{\bibinfo{volume}{87}},
  \bibinfo{pages}{49--58}, \doiprefix\url{10.1016/j.cjph.2023.09.015}
  (\bibinfo{year}{2024}).
\newblock \eprint{2303.01698}.

\bibitem{Gonzalez:2023rsd}
\bibinfo{author}{Gonz\'alez, E.}, \bibinfo{author}{Jusufi, K.},
  \bibinfo{author}{Leon, G.} \& \bibinfo{author}{Saridakis, E.~N.}
\newblock \bibinfo{journal}{\bibinfo{title}{{Observational constraints on
  Yukawa cosmology and connection with black hole shadows}}}.
\newblock {\emph{\JournalTitle{Phys. Dark Univ.}}}
  \textbf{\bibinfo{volume}{42}}, \bibinfo{pages}{101304},
  \doiprefix\url{10.1016/j.dark.2023.101304} (\bibinfo{year}{2023}).
\newblock \eprint{2305.14305}.

\bibitem{Sahoo:2023czj}
\bibinfo{author}{Sahoo, S.~K.}, \bibinfo{author}{Yadav, N.} \&
  \bibinfo{author}{Banerjee, I.}
\newblock \bibinfo{journal}{\bibinfo{title}{{Imprints of
  Einstein-Maxwell-dilaton-axion gravity in the observed shadows of Sgr A* and
  M87*}}}.
\newblock {\emph{\JournalTitle{Phys. Rev. D}}} \textbf{\bibinfo{volume}{109}},
  \bibinfo{pages}{044008}, \doiprefix\url{10.1103/PhysRevD.109.044008}
  (\bibinfo{year}{2024}).
\newblock \eprint{2305.14870}.

\bibitem{Nozari:2023flq}
\bibinfo{author}{Nozari, K.} \& \bibinfo{author}{Saghafi, S.}
\newblock \bibinfo{journal}{\bibinfo{title}{{Asymptotically locally flat and
  AdS higher-dimensional black holes of
  Einstein\textendash{}Horndeski\textendash{}Maxwell gravity in the light of
  EHT observations: shadow behavior and deflection angle}}}.
\newblock {\emph{\JournalTitle{Eur. Phys. J. C}}}
  \textbf{\bibinfo{volume}{83}}, \bibinfo{pages}{588},
  \doiprefix\url{10.1140/epjc/s10052-023-11755-w} (\bibinfo{year}{2023}).
\newblock \eprint{2305.17237}.

\bibitem{Uniyal:2023ahv}
\bibinfo{author}{Uniyal, A.}, \bibinfo{author}{Chakrabarti, S.},
  \bibinfo{author}{Fathi, M.} \& \bibinfo{author}{\"Ovg\"un, A.}
\newblock \bibinfo{journal}{\bibinfo{title}{{Observational signatures: Shadow
  cast by the effective metric of photons for black holes with rational
  non-linear electrodynamics}}}.
\newblock {\emph{\JournalTitle{Annals Phys.}}} \textbf{\bibinfo{volume}{462}},
  \bibinfo{pages}{169614}, \doiprefix\url{10.1016/j.aop.2024.169614}
  (\bibinfo{year}{2024}).
\newblock \eprint{2309.13680}.

\bibitem{Filho:2023ycx}
\bibinfo{author}{Filho, A. A.~A.}, \bibinfo{author}{Reis, J. A. A.~S.} \&
  \bibinfo{author}{Hassanabadi, H.}
\newblock \bibinfo{journal}{\bibinfo{title}{{Exploring antisymmetric tensor
  effects on black hole shadows and quasinormal frequencies}}}.
\newblock {\emph{\JournalTitle{JCAP}}} \textbf{\bibinfo{volume}{05}},
  \bibinfo{pages}{029}, \doiprefix\url{10.1088/1475-7516/2024/05/029}
  (\bibinfo{year}{2024}).
\newblock \eprint{2309.15778}.

\bibitem{EventHorizonTelescope:2022xqj}
\bibinfo{author}{Akiyama, K.} \emph{et~al.}
\newblock \bibinfo{journal}{\bibinfo{title}{{First Sagittarius A* Event Horizon
  Telescope Results. VI. Testing the Black Hole Metric}}}.
\newblock {\emph{\JournalTitle{Astrophys. J. Lett.}}}
  \textbf{\bibinfo{volume}{930}}, \bibinfo{pages}{L17},
  \doiprefix\url{10.3847/2041-8213/ac6756} (\bibinfo{year}{2022}).
\newblock \eprint{2311.09484}.

\bibitem{Raza:2023vkn}
\bibinfo{author}{Raza, M.~A.} \emph{et~al.}
\newblock \bibinfo{journal}{\bibinfo{title}{{Shadow of novel rotating black
  hole in GR coupled to nonlinear electrodynamics and constraints from EHT
  results}}}.
\newblock {\emph{\JournalTitle{Phys. Dark Univ.}}}
  \textbf{\bibinfo{volume}{44}}, \bibinfo{pages}{101488},
  \doiprefix\url{10.1016/j.dark.2024.101488} (\bibinfo{year}{2024}).
\newblock \eprint{2311.15784}.

\bibitem{Hoshimov:2023tlz}
\bibinfo{author}{Hoshimov, H.}, \bibinfo{author}{Yunusov, O.},
  \bibinfo{author}{Atamurotov, F.}, \bibinfo{author}{Jamil, M.} \&
  \bibinfo{author}{Abdujabbarov, A.}
\newblock \bibinfo{journal}{\bibinfo{title}{{Weak gravitational lensing and
  shadow of a GUP-modified Schwarzschild black hole in the presence of
  plasma}}}.
\newblock {\emph{\JournalTitle{Phys. Dark Univ.}}}
  \textbf{\bibinfo{volume}{43}}, \bibinfo{pages}{101392},
  \doiprefix\url{10.1016/j.dark.2023.101392} (\bibinfo{year}{2024}).
\newblock \eprint{2312.10678}.

\bibitem{Chakhchi:2024tzo}
\bibinfo{author}{Chakhchi, L.}, \bibinfo{author}{El~Moumni, H.} \&
  \bibinfo{author}{Masmar, K.}
\newblock \bibinfo{journal}{\bibinfo{title}{{Signatures of the accelerating
  black holes with a cosmological constant from the Sgr A\ensuremath{\star} and
  M87\ensuremath{\star} shadow prospects}}}.
\newblock {\emph{\JournalTitle{Phys. Dark Univ.}}}
  \textbf{\bibinfo{volume}{44}}, \bibinfo{pages}{101501},
  \doiprefix\url{10.1016/j.dark.2024.101501} (\bibinfo{year}{2024}).
\newblock \eprint{2403.09756}.

\bibitem{Liu:2024lve}
\bibinfo{author}{Liu, W.}, \bibinfo{author}{Wu, D.} \& \bibinfo{author}{Wang,
  J.}
\newblock \bibinfo{title}{{Shadow of slowly rotating Kalb-Ramond black holes}}
  (\bibinfo{year}{2024}).
\newblock \eprint{2407.07416}.

\bibitem{Liu:2024lbi}
\bibinfo{author}{Liu, W.}, \bibinfo{author}{Wu, D.}, \bibinfo{author}{Fang,
  X.}, \bibinfo{author}{Jing, J.} \& \bibinfo{author}{Wang, J.}
\newblock \bibinfo{journal}{\bibinfo{title}{{Kerr-MOG-(A)dS black hole and its
  shadow in scalar-tensor-vector gravity theory}}}.
\newblock {\emph{\JournalTitle{JCAP}}} \textbf{\bibinfo{volume}{08}},
  \bibinfo{pages}{035}, \doiprefix\url{10.1088/1475-7516/2024/08/035}
  (\bibinfo{year}{2024}).
\newblock \eprint{2406.00579}.

\bibitem{Joshi:2020tlq}
\bibinfo{author}{Joshi, A.~B.}, \bibinfo{author}{Dey, D.},
  \bibinfo{author}{Joshi, P.~S.} \& \bibinfo{author}{Bambhaniya, P.}
\newblock \bibinfo{journal}{\bibinfo{title}{{Shadow of a Naked Singularity
  without Photon Sphere}}}.
\newblock {\emph{\JournalTitle{Phys. Rev. D}}} \textbf{\bibinfo{volume}{102}},
  \bibinfo{pages}{024022}, \doiprefix\url{10.1103/PhysRevD.102.024022}
  (\bibinfo{year}{2020}).
\newblock \eprint{2004.06525}.

\bibitem{Dey:2020bgo}
\bibinfo{author}{Dey, D.}, \bibinfo{author}{Shaikh, R.} \&
  \bibinfo{author}{Joshi, P.~S.}
\newblock \bibinfo{journal}{\bibinfo{title}{{Shadow of nulllike and timelike
  naked singularities without photon spheres}}}.
\newblock {\emph{\JournalTitle{Phys. Rev. D}}} \textbf{\bibinfo{volume}{103}},
  \bibinfo{pages}{024015}, \doiprefix\url{10.1103/PhysRevD.103.024015}
  (\bibinfo{year}{2021}).
\newblock \eprint{2009.07487}.

\bibitem{Vagnozzi:2022moj}
\bibinfo{author}{Vagnozzi, S.} \emph{et~al.}
\newblock \bibinfo{journal}{\bibinfo{title}{{Horizon-scale tests of gravity
  theories and fundamental physics from the Event Horizon Telescope image of
  Sagittarius A}}}.
\newblock {\emph{\JournalTitle{Class. Quant. Grav.}}}
  \textbf{\bibinfo{volume}{40}}, \bibinfo{pages}{165007},
  \doiprefix\url{10.1088/1361-6382/acd97b} (\bibinfo{year}{2023}).
\newblock \eprint{2205.07787}.

\bibitem{Bambi:2019tjh}
\bibinfo{author}{Bambi, C.}, \bibinfo{author}{Freese, K.},
  \bibinfo{author}{Vagnozzi, S.} \& \bibinfo{author}{Visinelli, L.}
\newblock \bibinfo{journal}{\bibinfo{title}{{Testing the rotational nature of
  the supermassive object M87* from the circularity and size of its first
  image}}}.
\newblock {\emph{\JournalTitle{Phys. Rev. D}}} \textbf{\bibinfo{volume}{100}},
  \bibinfo{pages}{044057}, \doiprefix\url{10.1103/PhysRevD.100.044057}
  (\bibinfo{year}{2019}).
\newblock \eprint{1904.12983}.

\bibitem{Nojiri:2022cah}
\bibinfo{author}{Nojiri, S.} \& \bibinfo{author}{Nashed, G. G.~L.}
\newblock \bibinfo{journal}{\bibinfo{title}{{Consistency between black hole and
  mimetic gravity \textendash{} Case of (2+1)-dimensional gravity}}}.
\newblock {\emph{\JournalTitle{Phys. Lett. B}}} \textbf{\bibinfo{volume}{830}},
  \bibinfo{pages}{137140}, \doiprefix\url{10.1016/j.physletb.2022.137140}
  (\bibinfo{year}{2022}).
\newblock \eprint{2202.03693}.

\bibitem{Nojiri:2024txy}
\bibinfo{author}{Nojiri, S.} \& \bibinfo{author}{Odintsov, S.~D.}
\newblock \bibinfo{title}{{Improving Mimetic Gravity with Non-trivial Scalar
  Potential: Cosmology, Black Holes, Shadow and Photon Sphere}}
  (\bibinfo{year}{2024}).
\newblock \eprint{2408.05668}.

\end{thebibliography}

\end{document}